\documentclass[epj,fleqn]{svjour}
\usepackage[english]{babel}%
\usepackage{amsmath,amssymb}%
\usepackage{leftidx}%
\usepackage{graphicx}%
\usepackage[normal]{caption}
\usepackage{grffile}%
\usepackage{ifthen}%
\usepackage[squaren]{SIunits}%
\usepackage{xcolor}%
\usepackage{hyperref}%
\usepackage{dcolumn} \usepackage{bm}
\usepackage[numbers,sort&compress]{natbib}
\usepackage{url}

\newcommand{\subref}[2]{(#2)} \newcommand{\subfloat}[1]{}

\newcommand{\cf}{{cf.\ }}                            
\newcommand{\Tdefinition}[1]{{#1}}                    
\newcommand{\eq}{eq.}                                     
\newcommand{\Eq}{Eq.}                                     
\newcommand{\eqs}{eqs.}                                   
\newcommand{\file}[1]{\nolinkurl{#1}}                           
\newcommand{\fig}{fig.}                                   
\newcommand{\Fig}{Fig.}                                   
\newcommand{\figs}{figs.}                                 
\newcommand{\ie}{{i.\,e.\ }}                         
\newcommand{\eg}{{e.\,g.\ }}                         

\newcommand{\tab}{tab.}                                   
\newcommand{\via}{{via}\ }                             

\newcommand{\treref}[2]{[notes p.~#1 %
  \ifthenelse{\equal{#2}{}}{} {eq.~(#2)}%
]}                                                  
\newcommand{\trerevision}[1]{#1}

\newcommand{\BesselIsymbol}{\mathrm{I}}
\newcommand{\BesselI}[2]{\BesselIsymbol_{#1}\left(#2\right)}
\newcommand{\BesselIpow}[3]{\BesselIsymbol_{#1}^{#3}\left(#2\right)}
\newcommand{\eaverage}[1]{\langle{#1}\rangle}             
\newcommand{\eqspace}{\;}                              
\newcommand{\expectation}[1]{\langle{#1}\rangle}          
\newcommand{\matsc}[3]{#1 
  \ifthenelse{\equal{#2}{}}{}{^{(#2)}}%
  \ifthenelse{\equal{#3}{}}{}{_{#3}}%
}%
\renewcommand{\vec}[1]{{\bm{#1}}}                  
\newcommand{\vecdl}[1]{{\tilde{\bm{#1}}}}                  
\newcommand{\vecsc}[3]{#1 
  \ifthenelse{\equal{#2}{}}{}{^{(#2)}}%
  \ifthenelse{\equal{#3}{}}{}{_{#3}}%
}%
\addunit{\molar}{M}                                       
\addunit{\calory}{cal}                                    

\newcommand{\dl}{\tilde}                       
\newcommand{\Gr}{r}                                       

\newcommand{\drm}{\mathrm{d}}                  
\newcommand{\e}{\mathrm{e}}                               
\newcommand{\kBT}{\mathrm{k_B}T}                          
\newcommand{\kronecker}[1]{\delta_{#1}}                   
\newcommand{\standarddeviation}[1]{{\sigma_{#1}}}         
\newcommand{\timesspace}{\ }                   
\newcommand{\Ttime}{t}                                       

\newcommand{\msd}[1]{\mathrm{MSD}(#1)}

\newcommand{\beadradius}{a}                               
\newcommand{\Ia}{\beadradius}                             
\newcommand{\axialdensityprofile}{\rho}                  
\newcommand{\Iboxsize}{L}                                 
\newcommand{\Ibinwidth}{{l_{\mathrm{b}}}}               
\newcommand{\Icontourlength}{{L_{\mathrm{c}}}}                           
\newcommand{\diffusionconstant}{D}                        
\newcommand{\ID}{\diffusionconstant}                      
\newcommand{\IDo}{\ID_{0}}                                
\newcommand{\IDcG}{\diffusionconstant_{\mathrm{G}}} 
\newcommand{\Ieps}{\varepsilon}                           
\newcommand{\Iepsdl}{\dl \Ieps}                           
\newcommand{\Iepsc}{\Ieps_{\mathrm{s}}}                  
\newcommand{\Iepscdl}{\Iepsdl_{\mathrm{s}}}                  
\newcommand{\IepsG}{\Ieps_{\mathrm{G}}}              
\newcommand{\IepsGdl}{\Iepsdl_{\mathrm{G}}}                  
\newcommand{\Iepsscale}{\theta}                              
\newcommand{\Iextension}{x}                               
\newcommand{\IF}{F}                                       
\newcommand{\IFeq}{\IF_{\mathrm{eq}}}                     
\newcommand{\IFfriction}{\IF_{\mathrm{fr}}}               
\newcommand{\IFp}{\IF_{\mathrm{p}}}                       
\newcommand{\Ifrictionexponent}{\gamma}                   
\newcommand{\Ifrictioncoefficient}{\Gamma}
\newcommand{\Ifrictioncoefficientwork}{\tilde\Gamma}
\newcommand{\IlG}{l_\mathrm{G}}                           
\newcommand{\ImobilityO}{{\mu_0}}                              
\newcommand{\IN}{N}                                       
\newcommand{\ING}{\IN_{\mathrm{G}}}                       
\newcommand{\INGred}{\IN_{\mathrm{G}}'}                       
\newcommand{\In}{i}                                       
\newcommand{\Inc}{\In_{\mathrm{c}}}                         
\newcommand{\Ipotential}{U}                               
\newcommand{\Ipotentialdl}{\dl \Ipotential}               
\newcommand{\Ipotentialbond}{\Ipotential_{\mathrm{b}}}      
\newcommand{\Ipotentialring}{\Ipotential_{\mathrm{r}}}      
\newcommand{\IpotentialbondK}{\kappa}                      
\newcommand{\Ipotentialtrap}{\Ipotential_{\mathrm{tr}}}   
\newcommand{\IpotentialtrapK}{\kappa_{\mathrm{tr}}}            
\newcommand{\Ipotentialtrapn}{\In_{\mathrm{tr}}}            
\newcommand{\Ipotentialtrapr}{\vecsc{\vec R}{}{}} 
\newcommand{\IpotentialtrapPos}{{\vecsc{R}{}{x}}}            
\newcommand{\IpotentialtrapPosmin}{{\vecsc{R}{\mathrm{min}}{x}}}            
\newcommand{\IpotentialtrapPosmax}{{\vecsc{R}{\mathrm{max}}{x}}}            
\newcommand{\IpotentialtrapPosyz}{{\vecsc{R}{}{y/z}}}            
\newcommand{\IpotentialLJ}{\Ipotential_{\mathrm{LJ}}}       
\newcommand{\Ipotentialperiodic}{\Ipotential_{\mathrm{p}}}  
\newcommand{\Irandomforce}{f}                             
\newcommand{\Irandomforcedl}{\dl \Irandomforce}           
\newcommand{\Ir}{\Gr}                                     
\newcommand{\Irnp}{\hat{\vec{\Gr}}}                             
\newcommand{\IrcG}{\vecsc{\vec\Ir}{\text{G}}{}}          
\newcommand{\IxcG}{\vecsc{\Ir}{\mathrm{G}}{x}}           
\newcommand{\Iri}{{\vecsc{\vec\Ir}{i}{}}}                 
\newcommand{\Iridl}{{\vecsc{\vecdl\Ir}{i}{}}}            
\newcommand{\Irj}{{\vecsc{\vec\Ir}{j}{}}}                 
\newcommand{\It}{\Ttime}                                  
\newcommand{\Itdl}{\dl \It}                               
\newcommand{\Idt}{\Delta\It}                              
\newcommand{\Idtdl}{\Delta\Itdl}                          
\newcommand{\Itd}{\tau}                                   
\newcommand{\Iv}{v}                                       
\newcommand{\Ivdl}{\dl \Iv}                               
\newcommand{\Ieta}{{\eta}}                          
\newcommand{\Ietao}{{\eta_0}}                       
\newcommand{\IetaG}{{\eta_{\mathrm{G}}}}            
\newcommand{\IW}{W}                                       
\newcommand{\IWeq}{\IW_{\mathrm{eq}}}                     
\newcommand{\IWdiss}{\Delta \IW}                          

\DeclareMathOperator{\dist}{\mathcal{D}}                         

\title{Conformational dynamics and internal friction in homo-polymer
  globules: equilibrium vs. non-equilibrium simulations }

\titlerunning{Conformational dynamics and internal friction }

\abstract{ We study the conformational dynamics within homo-polymer
  globules by solvent-implicit Brownian dynamics simulations.
  A strong dependence of the internal chain dynamics on the
  Lennard-Jones cohesion strength~$\Ieps$ and the globule size~$\ING$
  is observed. We find two distinct dynamical regimes: a liquid-like
  regime (for $\Ieps<\Iepsc$) with fast internal dynamics and a
  solid-like regime (for $\Ieps>\Iepsc$) with slow internal dynamics.
  The cohesion strength $\Iepsc$ of this freezing transition depends
  on $\ING$.
  Equilibrium simulations, where we investigate the diffusional chain
  dynamics within the globule, are compared with non-equilibrium
  simulations, where we unfold the globule by pulling the chain ends
  with prescribed velocity (encompassing low enough velocities so that
  the linear-response, viscous regime is reached).
  From both simulation protocols we derive the internal viscosity
  within the globule.  In the liquid-like regime the internal friction
  increases continuously with $\Ieps$ and scales extensive in
  $\ING$. This suggests an internal friction scenario where the entire
  chain (or an extensive fraction thereof) takes part in
  conformational reorganization of the globular structure. }

\author{Thomas R. Einert\inst{1} \and Charles E. Sing\inst{2} \and
  Alfredo Alexander-Katz\inst{2} \and Roland R. Netz\inst{3}}
\institute{Physik Department, Technische Universit\"at M\"unchen,
  James-Franck-Stra\ss e, 85748 Garching, Germany \and Department of
  Materials Science and Engineering, Massachusetts Institute of
  Technology, Cambridge, MA 02139-4307, U.S.A \and Fachbereich Physik,
  Freie Universit\"at Berlin, Arnimallee 14, 14195 Berlin, Germany,
  \email{rnetz@physik.fu-berlin.de} } \date{\today}

\begin{document}
\onecolumn

\maketitle




\section{Introduction}
\label{sec:introduction}

Conformational dynamics of polymers play a crucial role in biological
systems. For example, pore-translocation of polymers such as RNA
requires bond breaking and drastic conformational rearrangements to
accommodate the geometrical constraints imposed by a
pore~\cite{Chuang2001,Bundschuh2005,Luo2009}.  Mechanical unfolding of
biopolymers in force spectroscopy experiments induces profound
conformational changes~\cite{Gunari2007}, which typically involve
dissipative effects~\cite{Rief1997,Staple2008}.  RNA sequences known
as riboswitches experience conformational changes upon binding of
small metabolites~\cite{Mandal2004}. All of these transitions
necessitate spatial rearrangements of the molecules and in certain
cases require chain reptation within a collapsed
region~\cite{Gennes1971}. The time scale on which these changes happen
is influenced by the medium, particularly the solvent viscosity, and
by the polymer itself, through internal interactions. Therefore, if
one desires to understand the dynamics of processes such as protein
folding~\cite{Jas2001,Pabit2004,Graeter2008,Sagnella2000}, packing of
DNA in the
chromosome~\cite{Grosberg1988,Grosberg1993,Lieberman-Aiden2009,Poirier2001},
polymer collapse~\cite{Frisch2002,Denesyuk2009,Abrams2002}, or
adsorption~\cite{Serr2010,Celestini2004,Chuang2000}, all dissipation
and viscous effects have to be considered~\cite{Barsegov2008}.

While the viscosity of the medium certainly influences polymer chain
dynamics, there are a number of effects that together give rise to
what is called internal friction. Local interactions such as
conformational transitions of backbone bonds and dihedral
angles~\cite{Jas2001,Khatri2007}, entanglement effects and excluded
volume interactions in polymer systems, degrees of freedom orthogonal
to the reaction coordinate~\cite{Hansen2010}, and the breakage and
reformation of cohesive
bonds~\cite{Filippov2004,Pabit2004,Murayama2007} all lead to
dissipation and thus increase the internal friction.  For globular
homopolymers, proteins in the molten globule
phase~\cite{Finkelstein2004}, and disordered intermediates during
protein folding~\cite{Gerber2007}, these effects can be conceptualized
as roughness on a hypothetical free energy landscape, corresponding to
many competing and intermediate states~\cite{Zwanzig1988,Hyeon2003},
leading directly to the idea of an effective internal viscosity
landscape~\cite{Alexander-Katz2009,Hinczewski2010,Best2010}.  There
have been a large number of coarse-grained simulation studies on the
force-induced unfolding of proteins~\cite{Klimov1999}, globular
polymers~\cite{Yoshinaga2005,Morrison2007,Celestini2004,Frisch2002,Braun2004a},
and the diffusion of knots along a stretched
chain~\cite{Metzler2006,Vologodskii2006,Huang2007}. Cohesive
interactions between polymer monomers have been shown to lead to a
phase transition from a liquid-like to a solid-like globule for long
enough
chains~\cite{Sing2010,Parsons2006,Parsons2006a,Paul2007,Rampf2005,Zhou1996,Taylor2009,Rostiashvili2001,Liang2000}.

In this paper we study internal friction in two model systems, which
both can be realized experimentally. We perform solvent-implicit
Brownian dynamics simulations on a homopolymer. Attractive
interactions are modeled with a Lennard-Jones potential, where the
cohesive strength~$\Ieps$ is varied.  First, we study the diffusion of
a globule, which forms from a polymer chain held at constant extension
smaller than the contour length. This simulation is conducted under
equilibrium conditions and no external forces are applied.
\trerevision{For $\Ieps<\Iepsc$, where $\Iepsc$ depends on the globule
  size, pronounced diffusion of the globule relative to the linker
  chain section is observed and characterized by the globule
  diffusivity~$\IDcG$.  $\IDcG$ is a direct measure of the internal
  globule viscosity, since motion of the globule relative to the
  linker chain requires internal rearrangements. } We observe that the
diffusion constant is proportional to the reciprocal globule size
$\IDcG\sim\ING^{-1}$ and shows a marked dependence on the cohesive
strength~$\Ieps$. For $\Ieps>\Iepsc$, no diffusion is observed on the
simulation time scales and the globule is stuck in a single
conformation. This reflects the change of the internal dynamics going
from liquid-like ($\Ieps<\Iepsc$) to solid-like
($\Ieps>\Iepsc$). Support for this interpretation is given by our
second set of simulations, where we measure the dissipated work during
stretching and unraveling the globule with a prescribed finite
velocity.  This is the same setup as studied by us before, but here
with significantly longer chains and slower pulling
velocities~\cite{Alexander-Katz2009}.  Like in our equilibrium
simulations, we observe two different regimes. For large
$\Ieps>\Iepsc$, the force extension curves are characterized by
pronounced fluctuations, which are absent for $\Ieps<\Iepsc$.
Pulling decreases the number of monomers inside the globule, causing
the fluctuations to vanish once $\ING$ is below a certain threshold.
\trerevision{Therefore, reducing the number of monomers~$\ING$ in the
  globule by stretching the chain, drives the system from the solid
  into the liquid state, similar to the reduction of the glass
  temperature of polymers close to
  interfaces\cite{Pablo1,Pablo2,Baschnagel1,Baschnagel2}.}  In the
liquid state we perform extensive simulations and show that the
internal viscosity is extensive, meaning that the dissipated work per
monomer that is pulled out from the globule scales linearly with the
globular monomer number $\ING$ and the pulling velocity~$\Iv$, similar
to our findings from the equilibrium simulations.  For the chain
dynamics this means that in the liquid-like regime, a model based on
local viscous friction is valid, but that the viscosity depends on the
size of a globule.  Clearly, internal friction effects dramatically
influences the time scales of chain dynamics in globules.  In the
solid-like regime, no definite conclusion on the conformational
dynamics is possible from the simulations we performed.

\section{Brownian dynamics simulations}
\label{sec:simulation-method}


We model the homopolymer by $\IN$ freely jointed beads of
radius~$\Ia$, interacting with a
potential~$\Ipotential(\{\vecsc{\vec\Ir}{\IN}{}\})$, which depends on
the set of positions of all beads $\{\vecsc{\vec\Ir}{\IN}{}\}$. The
position $\Iri$ of the $i$\textsuperscript{th} bead obeys the
overdamped Langevin equation~\cite{Doi1999,Coffey2005}
\begin{equation}
  \label{eq:1}
  \frac{\partial \Iri}{\partial \It} = - \ImobilityO \vec{\nabla}_\Iri
  \Ipotential(\{\vecsc{\vec\Ir}{\IN}{}\}) + \ImobilityO \vecsc{\vec\Irandomforce}{i}{}(\It)
  \eqspace,
\end{equation}
where $\ImobilityO = 1/(6\pi\Ietao a)$ is the Stokes mobility of a
sphere with radius~$\Ia$ in a solvent with viscosity~$\Ietao$.
$\vecsc{\vec\Irandomforce}{i}{}(\It)$ is the random force acting on
the $i$\textsuperscript{th} bead.  The components of the random
force~$\vecsc{\Irandomforce}{i}{\alpha}(\It)$, $\alpha = x,\,y,\,z$,
satisfy the Stokes-Einstein relation
\begin{equation}
  \eaverage{\vecsc\Irandomforce{i}{\alpha}(\It)\timesspace\vecsc\Irandomforce{j}{\beta} (\It')} = 
  \frac{2 \kBT}{\ImobilityO}\timesspace \kronecker{ij}\kronecker{\alpha\beta}\delta(\It-\It')
  \label{eq:2}
\end{equation}
and are unbiased $\eaverage{\vecsc\Irandomforce{i}{\alpha}(\It)} = 0$.
Hydrodynamic interactions are neglected, since we are interested in
the internal friction of the polymer caused by the monomer
interactions.
We express energies in units of the thermal energy $\kBT$, lengths in
units of the bead radius $\Ia$, and times in units of $\Itd = \Ia^2 /
(\ImobilityO\kBT)$, which is the characteristic diffusion time of a
single bead. Using the dimensionless quantities $\Itdl = \It/\Itd$,
$\vecdl\Ir = \vec\Ir/\Ia$, $\vecdl\Irandomforce =
\vec\Irandomforce/(\kBT/\Ia)$, and $\Ipotentialdl =
\Ipotential/(\kBT)$, \eq~\eqref{eq:1} reads
\begin{equation}
  \label{eq:3}
  \frac{\partial \Iridl}{\partial \Itdl} = - \vec{\nabla}_{\Iridl}
  \Ipotentialdl(\{\vecsc{\vecdl\Ir}{\IN}{}\}) + \vecsc{\vecdl\Irandomforce}{i}{}(\Itdl)
  \eqspace.
\end{equation}

For the Brownian dynamics simulation we use a discretized version of
\eq~\eqref{eq:3} with a time step $\Idt = 0.0005\Itd$ and obtain the
coordinates at time $\It = n \Idt$~\cite{Ermak1978}
\begin{equation}
  \label{eq:4}
  \Iridl(n+1)=   \Iridl(n) + \left(- \vec{\nabla}_{\Iridl}
    \Ipotentialdl(\{\vecsc{\vecdl\Ir}{\IN}{}\}) + \vecsc{\vecdl\Irandomforce}{i}{}(n) \right)\Idtdl 
\end{equation}
and
\begin{equation}
  \label{eq:5}
  \eaverage{\vecsc\Irandomforcedl{i}{\alpha}(n)\timesspace\vecsc\Irandomforcedl{j}{\beta}
    (n')} = \frac{2}{\Idtdl}\timesspace \kronecker{ij}\kronecker{\alpha\beta}\kronecker{nn'}
  \eqspace. 
\end{equation}

\section{Diffusivity of a globule along a periodic chain: equilibrium
  simulations}
\label{sec:diff-glob-along-1}

In our preceding work we determined the internal viscosity~$\IetaG$ of
a homopolymeric globule by measuring the dissipated energy when
unfolding the globule by pulling apart the chain ends at finite
speed~\cite{Alexander-Katz2009}. We showed that for moderate cohesion
$\Ieps$ the internal friction is the dominant dissipative effect as
long as the majority of the monomers are part of the globule. However,
as the globule is unraveled, the size of the globule decreases and
more and more beads in the linker sections dissipate energy due to
solvent friction. These effects had to be subtracted in order to
obtain the internal globule friction.  Likewise, reaching the relevant
linear-response regime at low pulling velocities is subtle.  Here, we
introduce a novel system setup, where we perform equilibrium diffusion
simulations with a globule of constant size:
The entire polymer chain is held at constant extension smaller than
the contour length, so that a globule forms for high enough cohesive
energy.  \trerevision{Since we fix the position of the linker chain in
  space, as we explain further below, the diffusive motion of the
  globule necessitates internal rearrangements, so that the globular
  diffusivity relative to the linker chain is a measure of the globule
  internal friction.} Since the method works at equilibrium, the
linear viscous regime is automatically obtained.
By this method, the internal viscosity inside the globule manifests
itself as a macroscopic and experimentally observable quantity: the
diffusivity of the globule, $\IDcG\sim 1/\IetaG$.  This system could
be realized experimentally by a polymer held at constant extension in
an optical or magnetic tweezers setup~\cite{Gebhardt2010}.  In such an
experiment the cohesive force can be varied by changing the solution
conditions, whereas the size of the globule can be varied by changing
the trap distance.

\subsection{Model}
\label{sec:model-2}

\subsubsection{Description of the system}
\label{sec:description-system}
\begin{figure}
  \centering

  \includegraphics{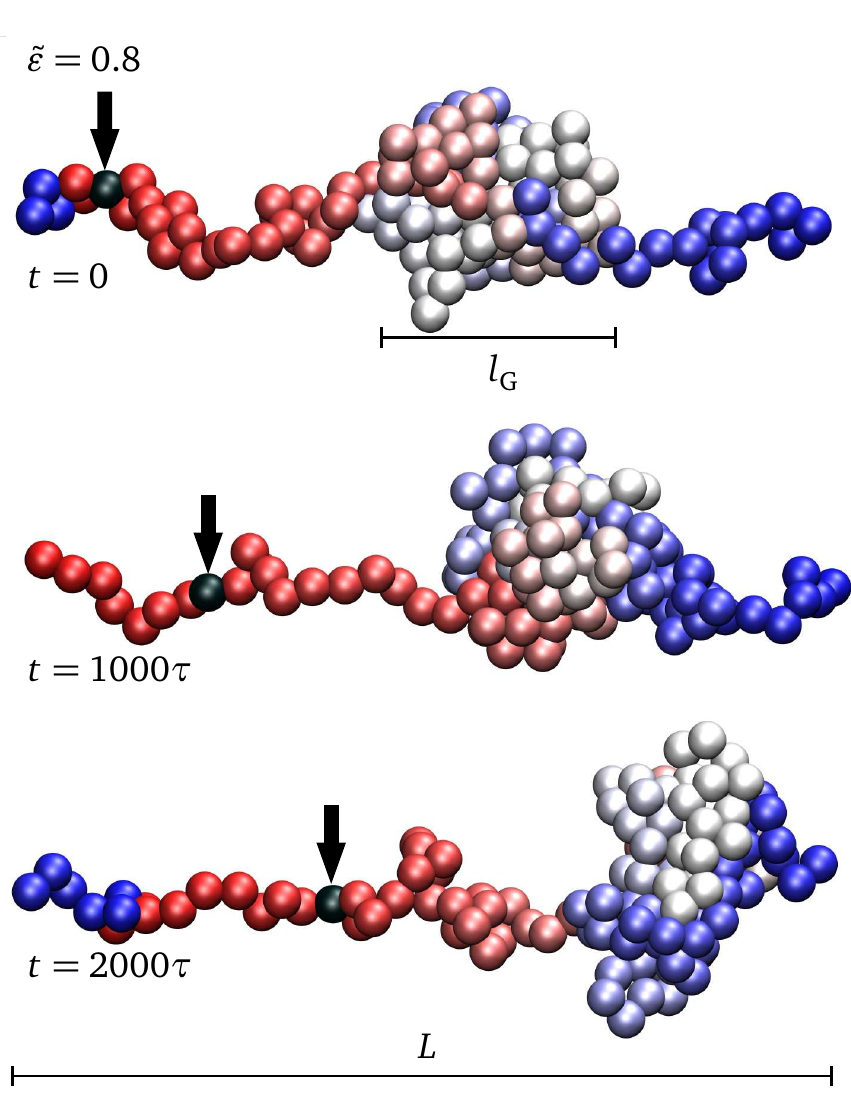}%
  \includegraphics{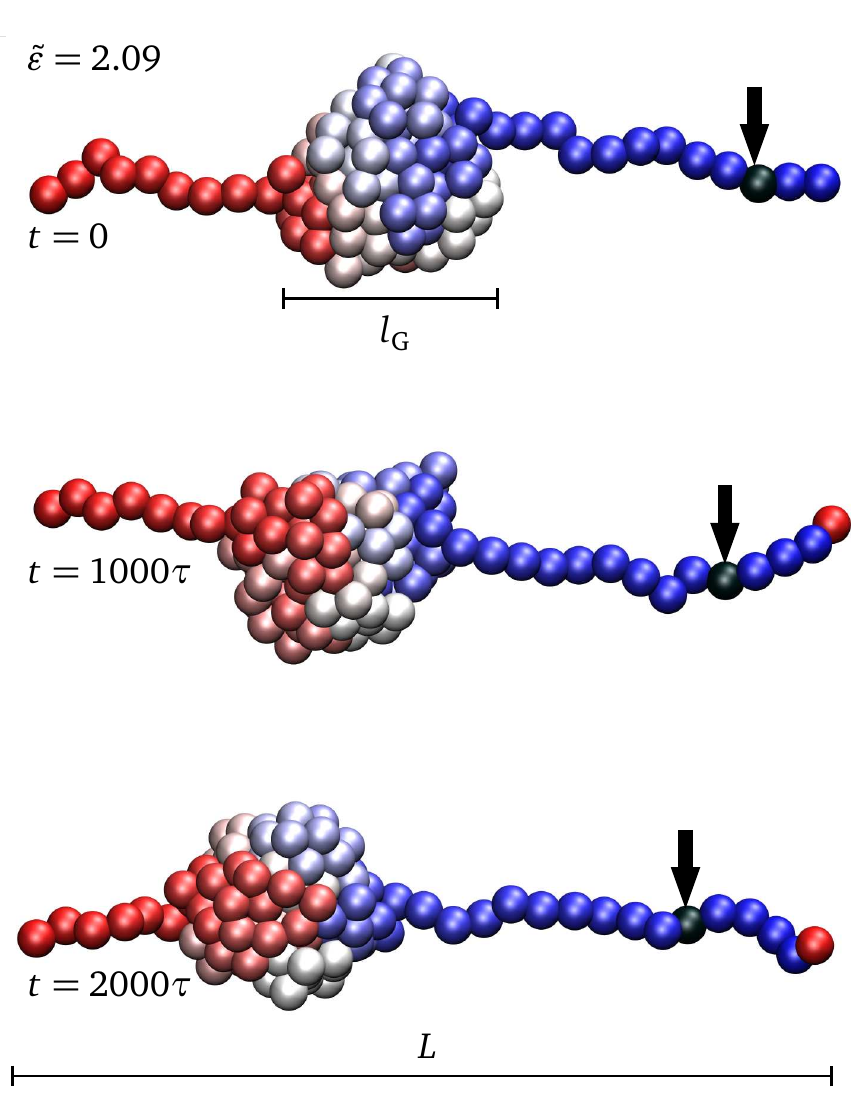}
  \caption{ The diffusion of a globule relative to the stretched
    polymer linker is simulated. The globule mobility, the number of
    monomers~$\ING$ inside the globule, the diameter of the globule
    $\IlG$, and the fluctuations of the linkers (i.e. the stretched
    part of the chain that is not part of the globule) depend on the
    cohesive strength~$\Ieps$, \eq~\eqref{eq:10}.  Snapshots at
    different times for $\Iepsdl = 0.8$ (left) and 2.09 (right) are
    shown.  To prevent motion of the linker chain section and thereby
    obtain directly the globule diffusivity relative to the chain, the
    monomer~$\Ipotentialtrapn$ in the middle of the linker chain
    (indicated by the arrow) is trapped by a harmonic potential,
    \eq~\eqref{eq:11}.  Periodic boundary conditions for a box of
    length~$\Iboxsize$ using the minimum image convention are employed
    to model an infinite polymer. The color coding indicates the
    running monomer index along the chain.  }
  \label{fig:1}
\end{figure}
We consider a polymer held at a fixed extension~$\Iboxsize = 50\Ia$ in
the $x$-direction, which is smaller than the contour length
$\Icontourlength = 2\Ia (\IN -1)$ of the polymer.  A globule will form
for large enough attractive Lennard-Jones interaction between the
monomers, see \fig~\ref{fig:1}.  To eliminate finite size effects, we
introduce periodic boundary conditions in $x$-direction, which are
implemented \via the minimum image convention~\cite{Allen1989}: The
components of the vector pointing from $\Iri$ to $\Irj$ are given by
\begin{subequations}\label{eq:6}
  \begin{gather}
    \vecsc{\Ir}{i,j}{x} =
    \dist(\vecsc{r}{i}{x},\vecsc{r}{j}{x};{\Iboxsize})\equiv
    ((\vecsc{r}{j}{x} -
    \vecsc{r}{i}{x} + 3\Iboxsize/2)\mod \Iboxsize) -\Iboxsize /2\\
    \vecsc{\Ir}{i,j}{y/z} = \vecsc{r}{j}{y/z} - \vecsc{r}{i}{y/z}
    \eqspace.
  \end{gather}
\end{subequations}
We use a box size $\Iboxsize = 50\Ia$ in all simulations.  The
potential energy has four contributions
\begin{equation}
  \Ipotential = \Ipotentialbond + \Ipotentialring + \IpotentialLJ + \Ipotentialtrap\label{eq:7}
  \eqspace.
\end{equation}
$\Ipotentialbond$ and $\Ipotentialring$ are the bond potentials acting
between neighboring monomers.  The backbone bonds are modeled by
harmonic potentials
\begin{equation}
  \Ipotentialbond = \frac{\IpotentialbondK}{2}
  \sum_{i=1}^{\IN-1}(\vecsc{\Ir}{i,i+1}{} - 2\Ia)^2\label{eq:8}
  \eqspace,
\end{equation}
with $\IpotentialbondK = 200\kBT/\Ia^2$ and $\vecsc{\Ir}{i,j}{} = |
{\bf r}^{(i,j)}|$, see \eq~\eqref{eq:6}.  As periodic boundary
conditions are employed, the polymer forms a closed ring, which is
achieved by connecting the first and last monomer by
\begin{equation}
  \label{eq:9}
  \Ipotentialring = \frac{\IpotentialbondK}{2} (\vecsc{\Ir}{\IN,1}{} -
  2\Ia)^2
  \eqspace.
\end{equation}
The monomer cohesion and excluded volume interactions are modeled with
a Lennard-Jones potential
\begin{equation}
  \IpotentialLJ = \Ieps
  \sum_{i=1}^\IN\sum_{j=1}^{i-1}\biggl(\left( \frac{2\Ia}{\vecsc{\Ir}{i,j}{}}\right)^{12} - 2 \left( \frac{2\Ia}{\vecsc{\Ir}{i,j}{}}\right)^{6}\biggr)\label{eq:10}
  \eqspace.
\end{equation}
$\Ieps = 0$ models an ideal phantom chain without excluded volume
interactions and attractive interactions between the monomers. For
$\Ieps>0$ the first term in \eq~\eqref{eq:10} accounts for the
repulsive excluded volume interaction at short separations, whereas
the second term is responsible for cohesion which reflects hydrophobic
attraction between monomers in a solvent-implicit fashion.
\trerevision{For $0\leq\Ieps<\IepsG\approx 0.5 $ the polymer is in the
  swollen state and no globule exists. Increasing $\Ieps$ above
  $\IepsG$ causes the polymer to collapse and a globule forms.  The
  static globule behavior from our simulations agrees with previous
  work~\cite{Alexander-Katz2006,Alexander-Katz2009,Sing2010,Parsons2006,Parsons2006a,Paul2007,Rampf2005,Zhou1996,Taylor2009,Rostiashvili2001,Liang2000,Frisch2002}.
  The value of the cohesive strength at the globule transition,
  $\IepsG$, depends on the system size~\cite{Parsons2006}. }For even
larger values $\Ieps>\Iepsc$ a solid phase appears.

As we are interested in the relative motion of the globule with
respect to the rest of the chain, we prevent the linker chain from
moving by an external trapping potential acting on one monomer,
$\Ipotentialtrap$, which can be viewed as the effect of an \eg optical
tweezers.  The linker chain is the stretched part that does not belong
to the globule, \fig~\ref{fig:1}. We introduce a harmonic trap
potential
\begin{equation}
  \label{eq:11}
  \Ipotentialtrap = \frac{\IpotentialtrapK}{2}(\vecsc{\vec\Ir}{\Ipotentialtrapn}{} - \Ipotentialtrapr)^2
  \eqspace,
\end{equation}
which is located at $ \Ipotentialtrapr$ and acts on
bead~$\Ipotentialtrapn$ that is in the middle of the linker, with
stiffness $\IpotentialtrapK = 10\kBT/\Ia^2$.  The index of the trapped
bead~$\Ipotentialtrapn$ depends on the position of the globule and on
the set of beads that belong to the globule.  The exact definition of
the globule is described in the next section.  Due to globule motion,
the index of the bead in the middle of the linker might change from
$\Ipotentialtrapn$ at time $\It$ to $\Ipotentialtrapn'$ at some later
time~$\It'$.  If this happens, we update the $x$-coordinate of the
trap position to the new position $\IpotentialtrapPos(\It') =
\IpotentialtrapPos(\It) + 2\Ia\dist(\Ipotentialtrapn,
\Ipotentialtrapn';\IN)$, \eq~\eqref{eq:6}, and the trapping force then
acts on bead $\Ipotentialtrapn'$. The trapped beads in the snapshots
shown in \fig~\ref{fig:1} are indicated by arrows.

This setup allows us to study the diffusion of the globule relative to
the linker chain along the $x$-axis. The effect of using periodic
boundary conditions rather than long linkers to each side is
fourfold. First, the simulation is sped up as the system is
smaller. Second, the globule is stabilized as the configurational
space for the unfolded system is reduced since only fluctuations up to
a wavelength of the order of the simulation box are possible. Third,
motion of the globule as a whole without internal friction is
reduced. Fourth, knots may not form as the polymer forms a closed
ring.

\subsubsection{Definition of the globule}
\label{sec:definition-globule}
\begin{figure}
  \centering
  \includegraphics{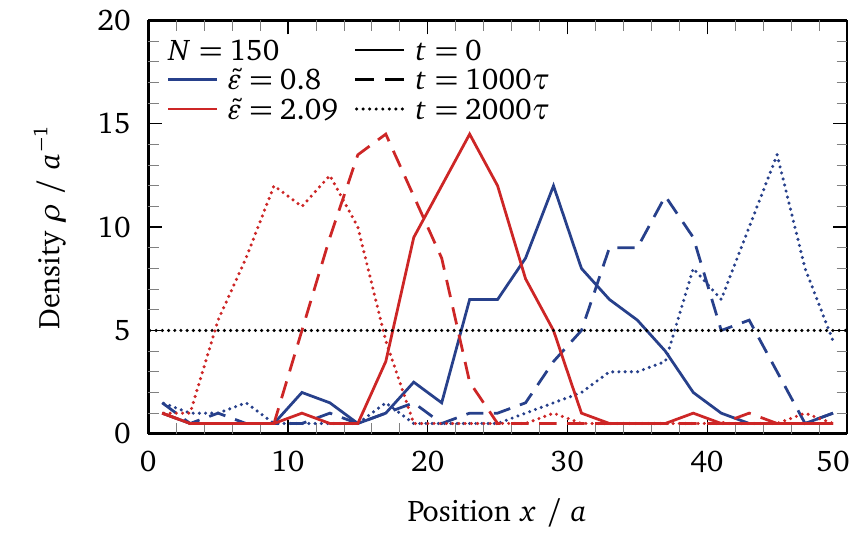}
  \caption{The density profile of the globule depends on the cohesive
    strength~$\Ieps$. The globule is defined as the region where the
    monomer density is $\axialdensityprofile>5/\Ia$ (indicated by the
    horizontal line).}
  \label{fig:2}
\end{figure}
We partition the $x$-axis into bins of width $\Ibinwidth = 2\Ia$ and
measure the monomer density~$\axialdensityprofile_k$ in the
$k$\textsuperscript{th} bin, $k = 1,\ldots,\Iboxsize/\Ibinwidth$.  The
globule is defined as the region, where the monomer density projected
on the $x$-axis fulfills the condition $\axialdensityprofile_k>5/\Ia$.
In \fig~\ref{fig:2} the density profiles of the snapshots in
\fig~\ref{fig:1} are shown.  If at the edge of the globule the density
profile is not monotonous and $\axialdensityprofile_k<5/\Ia$, but
$\axialdensityprofile_{k-1},\axialdensityprofile_{k+1}>5/\Ia$, we
account for such cases by adding these bins to the globule, too.  This
ensures that we end up with a list of bins that are connected.  For
small $\Iepsdl<1$ and small $\IN<150$ it is possible that --~according
to the above definition~-- more than one globule exists or that the
globule is smeared out over the complete simulation box, however we
will not consider simulations in which this occurs. Such complications
are never observed for $\Iepsdl>1$ and $\IN>150$.

\subsubsection{Definition of the index of the trapped bead}
\label{sec:diff-index-centr}

Let us first define the index of the bead $\In_{\mathrm{r}}$, which is
at the right edge of the globule, and the index of the
bead~$\In_{\mathrm{l}}$, which is at the left edge of the
globule. $\In_{\mathrm{r}}$ is obtained by picking a monomer inside
the globule and moving along the chain contour with increasing monomer
index. $\In_{\mathrm{r}}$ is the largest index that is still in a bin
belonging to the globule. We also check for loops, which leave the
globule and return again: If a loop occurs, we add all monomers of the
loop to the globule even if they lie in a bin outside the
globule. $\In_{\mathrm{l}}$ is defined in the same way yet by
decreasing the index.  The \Tdefinition{index of the central
  bead}~$\Inc$ in the middle of the globule and the number of
monomers~$\ING$ inside the globule are defined as $\Inc =
(\In_{\mathrm{l}} + \In_{\mathrm{r}})/2$, $\ING = \In_{\mathrm{r}} -
\In_{\mathrm{l}}+1$ if $\In_{\mathrm{l}}<\In_{\mathrm{r}}$ and $\Inc =
(\In_{\mathrm{l}}-\IN + \In_{\mathrm{r}})/2$, $\ING = \In_{\mathrm{l}}
- \In_{\mathrm{r}}+1$ if $\In_{\mathrm{l}}>\In_{\mathrm{r}}$.  The
index of the central bead~$\Inc$ yields the index of the trapped bead
via~$\Ipotentialtrapn = \Inc + \IN/2$, which therefore depends on the
motion of the globule.  We could have also trapped a fixed monomer for
the whole duration of the simulation. Our procedure (i.e. trapping a
bead that by construction is never part of the globule) allows maximal
dynamic freedom for the globule and thus improves equilibration of the
system.

\subsection{Results}
\label{sec:results}

\subsubsection{ Number of monomers inside the globule}
\label{sec:numb-monom-inside}

\begin{figure}
  \centering
  \includegraphics{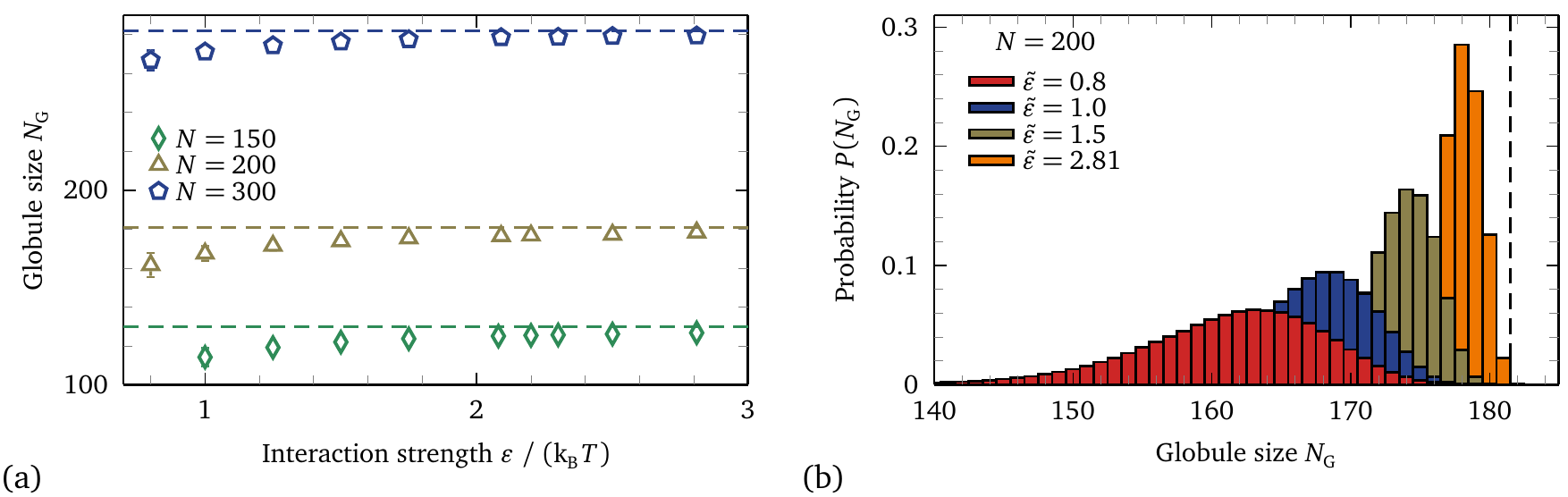}%
  \subfloat{\label{fig:3a}}%
  \subfloat{\label{fig:3b}}%
  \caption{ \subref{fig:3}a~The number of monomers $\ING$ inside the
    globule increases weakly with the interaction strength $\Ieps$
    until it finally levels off.  The limiting values $\ING^*$ are
    depicted by horizontal broken lines and are determined by
    \eq~\eqref{eq:12}, describing a spherical globule and tightly
    stretched linkers. Error bars denote the standard deviations of
    the distributions.  \subref{fig:3}b~Probability distribution of
    the number of monomers~$\ING$ inside the globule for chain
    length~$\IN = 200$ and different cohesive strengths $\Iepsdl =
    0.8,\,1,\,1.5,\,2.81$. For small $\Ieps$ considerable size
    fluctuations are observed, which decrease upon increasing
    $\Ieps$. Further, increasing $\Ieps$ raises the mean $\ING$ up to
    the limiting value $\ING^*$ given by \eq~\eqref{eq:12} (broken
    vertical line).  }
  \label{fig:3}
\end{figure}

The number of monomers~$\ING$ inside the globule depends on the chain
length~$\IN$ and on the cohesive strength $\Ieps$,
\fig~\ref{fig:3}a. Increasing $\Ieps$ raises $\ING$ only up to a
limiting value $\ING^*$, which is determined by the equation
\begin{equation}
  \label{eq:12}
  (\ING^*)^{1/3} + (\IN - \ING^*) = \Iboxsize/(2\Ia)
  \eqspace.
\end{equation}
\Eq~\eqref{eq:12} describes a close-packed spherical globule
consisting of $\ING^*$ monomers with a tightly stretched linker
consisting of $\IN-\ING^*$ monomers.  As the limiting value of our
simulations coincides with the predicted value of \eq~\eqref{eq:12},
as shown in \fig~\ref{fig:3}a, our definition of the globule is
justified.  For large $\Ieps$, the linkers are completely stretched,
as the energetic gain of a monomer joining the globule outweighs the
entropic loss of reducing the fluctuations of the linker.  For small
$\Ieps$, monomers are not tightly bound to the globule as can be seen
qualitatively from \fig~\ref{fig:1}. Therefore, size fluctuations of
the globule are more substantial for small cohesive strengths and
decrease upon increasing $\Ieps$, as shown in \fig~\ref{fig:3}b.

\subsubsection{Definition of the center of the globule}
\label{sec:diff-relat-cent}

In order to properly describe the motion of the polymer in a
simulation box with periodic boundary conditions in $x$-direction, we
introduce non-periodic polymer coordinates~$\{\vecsc{\Irnp}{\IN}{}\}$,
which are not restricted to the primary simulation cell. They are
defined recursively starting from the first monomer with
$\vecsc{\Irnp}{1}{} = \vecsc{\vec\Ir}{1}{}$ via
\begin{equation}
  \label{eq:13}
  \vecsc{\Irnp}{i+1}{} = \vecsc{\Irnp}{i}{} + \vecsc{\vec\Ir}{i,i+1}{}
  \eqspace,
\end{equation}
for $i>1$, where $\vecsc{\vec\Ir}{i,i+1}{}$ is given by
\eq~\eqref{eq:6}.  Since we are considering a ring-like polymer with
periodic boundary conditions in the $x$-direction, the center of mass
of the globule has to be defined carefully.
The center of the globule~$\IrcG$ is calculated by using the
non-periodic polymer coordinates, \eq~\eqref{eq:13}, according to
\begin{equation}
  \label{eq:14}
  \IrcG=\frac{1}{\ING} \sum_{i=\In_{\mathrm{l}}}^{\In_{\mathrm{r}}} \vecsc{ \Irnp}{i}{}
  \eqspace.
\end{equation}

Periodic boundary conditions may introduce jumps of the size of the
box $\Iboxsize$ in the trajectory~$\IrcG(\It)$. We remove those jumps
by connecting the value of any quantity $q(\It)$ at time~$\It$, which
is subject to periodic boundary conditions, with the value $q(\It -
\Delta\It)$ in the previous time step via the minimum image condition.

\subsubsection{Diffusivity of the globule}
\label{sec:diffusivity-globule}

\begin{figure}
  \centering
  \includegraphics{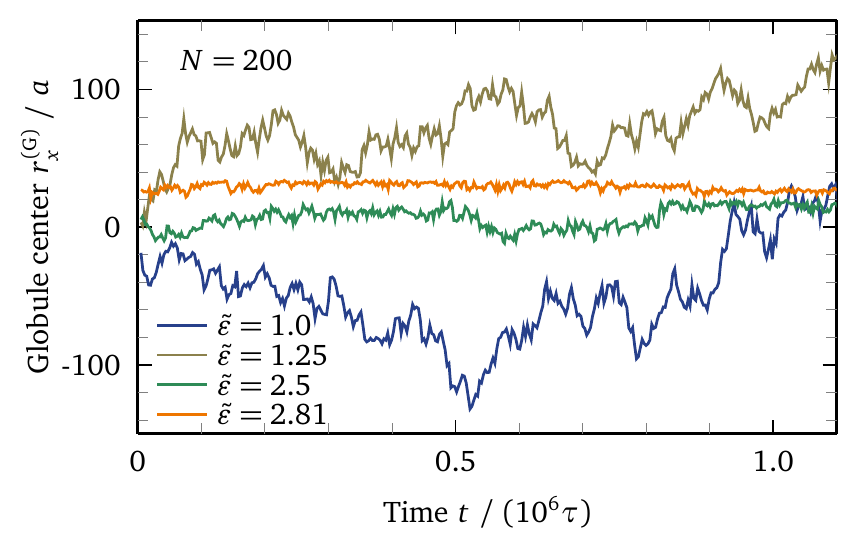}
  \caption{Trajectories of the $x$-coordinate of the center of the globule, $\protect\IxcG$,
  defined in \eq~\eqref{eq:14},  for
    $\IN = 200$ and various interaction strengths $\Iepsdl = 1,\,1.25,\,2.5,\,2.81$. 
    The smaller the attractive interaction between the monomers, the more
    mobile the globule is. If $\Ieps$ is large, the globule is frozen in a single
    conformation and does not move at all.
  }
  \label{fig:4}
\end{figure}

Our setup enables us to study the motion of the globule in space in a
fashion that is coupled to its internal conformational dynamics.
\trerevision{In other word, since we fix the linker chain in space,
  only internal chain rearrangements within the globule lead to
  diffusion of the globule in space.}  In \fig~\ref{fig:4}
trajectories of the $x$-coordinate $\IxcG$ of the center of the
globule, \eq~\eqref{eq:14}, are shown for $\IN = 200$ and various
cohesive strengths~$\Ieps$. \Fig~\ref{fig:4} demonstrates that the
diffusivity decreases with increasing $\Ieps$. As the monomers become
more cohesive, it is more difficult for the globule to rearrange
internally and hence to move. This effect is further accented by the
increase in globule size~$\ING$, see \fig~\ref{fig:3}, which
additionally decreases the mobility of the globule.

\begin{figure}
  \centering
  \includegraphics{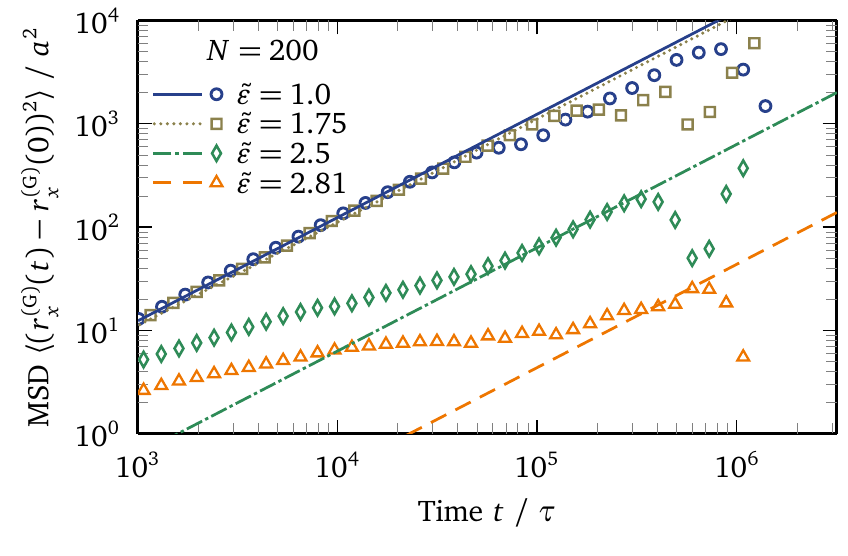}
  \caption{The mean squared displacement (MSD) of the center of the
    globule for $\IN = 200$ and $\Iepsdl = 1,\,1.75,\,2.5,\,2.81$ is
    calculated from the trajectories, \fig~\ref{fig:4}, using
    \eq~\eqref{eq:15}. Symbols denote the measured MSD from our
    simulations, lines the corresponding linear fits for
    $\msd{\It}\gtrsim10$. For $\Iepsdl<\Iepscdl $ with $ \Iepscdl
    \approx2.3$ for $\IN = 200$, the globule exhibits normal diffusion
    and the diffusivity~$\IDcG$ decreases as $\Ieps$ increases. For
    $\Iepsdl = 2.5$, normal diffusive behavior is obtained, whereas
    for $\Iepsdl=2.81$ the diffusion time scale is of the order of the
    simulation time and normal diffusive behavior is barely reached. }
  \label{fig:5}
\end{figure}
\begin{figure}
  \centering
  \includegraphics{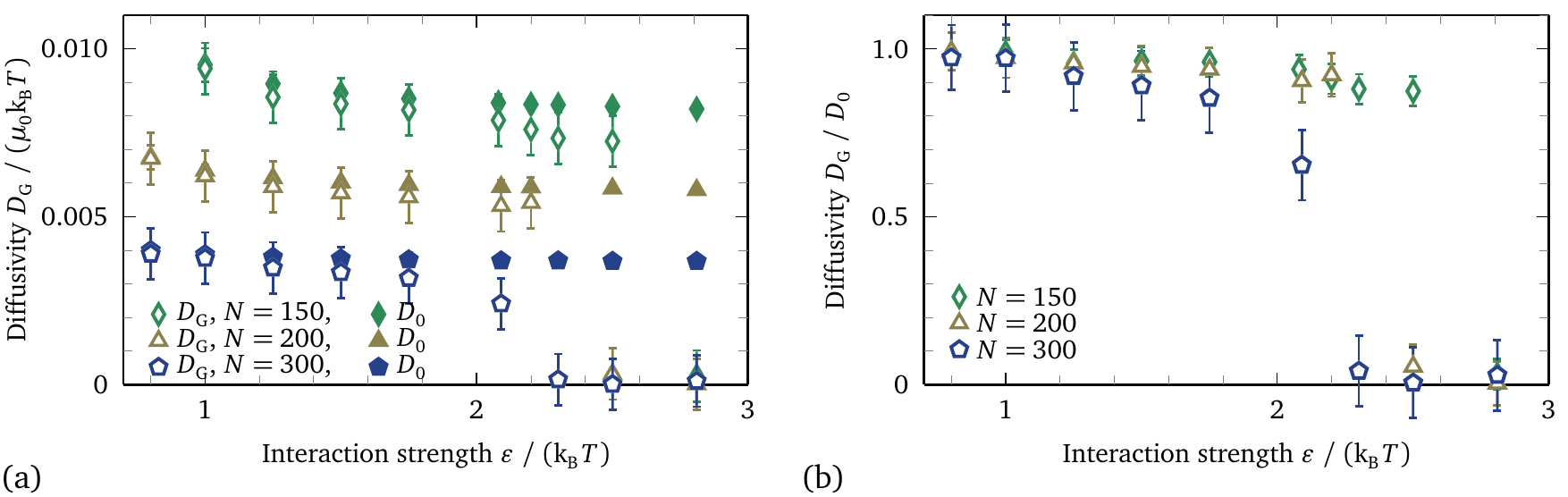}%
  \subfloat{\label{fig:6a}}%
  \subfloat{\label{fig:6b}}%
  \caption{\subref{fig:6}a~Diffusion constant~$\IDcG$ of the globule
    (open symbols) as obtained from linear fits to the MSD curves,
    \fig~\ref{fig:5}. $\IDcG$ decreases with increasing $\IN$ and
    $\Ieps$. At $\Ieps=\Iepsc$, with $\Iepsc$ depending on $\IN$, a
    transition of the internal dynamics occurs and the diffusion
    constant drops to zero. $\IDcG$ is compared to the ideal Rouse
    diffusion constant $\IDo$ (solid symbols) of a globule with
    $\INGred$ monomers, \eqs~\eqref{eq:16} and~\eqref{eq:17}, which
    can move freely and independent of the linkers.  (b)~Rescaling the
    actual diffusion constant by the ideal diffusion constant,
    $\IDcG/\IDo$, removes the $\IN$ dependence to a large extent in
    the liquid regime.  }
  \label{fig:6}
\end{figure}
\begin{figure}
  \centering
  \includegraphics{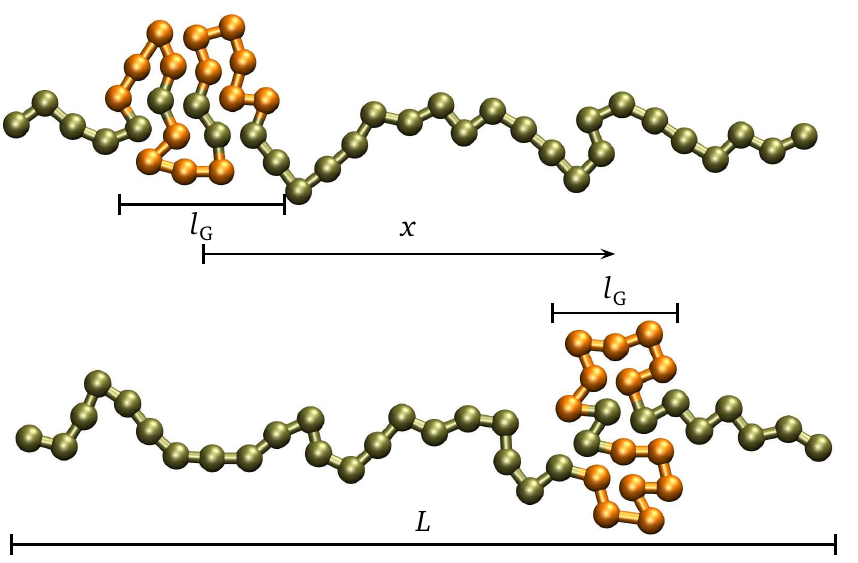}
  \caption{Illustration of the concept and calculation of the reduced
    number of monomers~$\INGred$ in the globule. For a displacement of
    the globule by a distance~$x$ only a reduced number of monomers
    have to move along, here $\INGred = 14$ (shown in orange). In
    order to obtain $\INGred$, the linker is continued through the
    globule and the monomers belonging to this internal linker are
    subtracted from $\ING$ yielding $\INGred$, see \eq~\eqref{eq:17}.}
  \label{fig:7}
\end{figure}

To quantify these observations, we calculate the mean squared
displacement (MSD) of $\IxcG$. For normal diffusive behavior one
expects the MSD to scale linearly with time and to be characterized by
the diffusion constant $\IDcG$
\begin{equation}
  \label{eq:15}
  \msd{\It} = \bigl\langle \bigl( \IxcG(\It) - \IxcG(0) \bigr)^2\bigr\rangle
  = \frac{1}{T}\int_0^T \bigl( \IxcG(\It+\It') - \IxcG(\It') \bigr)^2 \drm\It'
  = 2\IDcG \It
  \eqspace.
\end{equation}
MSD curves for $\IN= 200$ and various cohesive strengths are shown in
\fig~\ref{fig:5} on a double logarithmic plot. The MSD curves are
fitted with linear functions for $\msd{\It}\gtrsim10$ in order to
obtain the diffusion constant~$\IDcG$. For small $\Ieps<\Iepsc$,
normal diffusion is observed with $\IDcG$ decreasing as $\Ieps$
increases. However, as can be seen in \fig~\ref{fig:5} for $\IN = 200$
and $\Iepsdl = 2.5$, the normal diffusive regime only occurs at very
long time scales and for $\Iepsdl = 2.81$ is barely reached on the
time scales of our simulations.  We attribute this to a change of the
internal dynamics of the globule, which is slowed down with increasing
cohesion and is effectively suppressed for $\Ieps>\Iepsc$. One
remaining pathway for the globule to rearrange at very large cohesion
is to dissolve --~at least partly~-- and refold into a different
configuration.  As a consequence, the time scale characterizing the
internal dynamics should become comparable to the time scale on which
the globule dissolves.  This dissolution time is huge as it scales
exponentially with $\Ieps\ING$ and is beyond our simulation time. For
that reason we observe only stuck globules for large cohesive
strengths, which remain in a single conformation.  The fitted
diffusion constants are shown in \fig~\ref{fig:6} and contrasted with
an idealized limit, where internal friction is absent and the globule
and the linker move independently.  The diffusivity in this limit is
given by the Rouse diffusion constant of $\INGred$
monomers~\cite{Doi1999}
\begin{equation}
  \label{eq:16}
  \IDo = \frac{\ImobilityO\kBT}{\INGred}
  \eqspace.
\end{equation}
$\INGred\leq\ING$ is the reduced number of monomers within the globule
that actually have to comove when the globule is displaced by some
distance $x$.  $\INGred$ can be estimated by continuing the linker
through the globule and subtracting the number of monomers, which
belong to this internal linker section, from $\ING$, see
\fig~\ref{fig:7} for an illustration,
\begin{equation}
  \INGred
  = \ING - (\IN - \ING)\frac{\IlG}{\Iboxsize-\IlG}
  \label{eq:17}
  \eqspace.
\end{equation}
In \fig~\ref{fig:6}a a pronounced dependence of the diffusion constant
$\IDcG$ (open symbols) on the globule size $\ING$ is observed.  The
diffusivity decreases as the internal interactions $\Ieps$
increase. This is due to the coupling of the internal dynamics of the
globule to the overall motion of the globule, because, as explained
before, in our simulation setup the linker is fixed and the globule
can only move via internal conformational chain reorganization.
Slowing down the internal dynamics by increasing $\Ieps$ thus reduces
the mobility of the globule.  For small $\Ieps$ one observes
$\IDcG\approx\IDo$ (filled symbols), implying that internal friction
is unimportant. Conversely, increasing $\Ieps$ causes increasing
deviations between the mobility of the globule and the ideal system,
until finally $\IDcG$ drops to zero as the liquid-solid transition is
crossed.
\trerevision{For $\Ieps>\Iepsc$ the globule is in a frozen state with
  strongly suppressed internal dynamics. Since the linkers are
  trapped, this also impedes the motion of the globule as a whole and
  leads to a vanishing diffusion constant.}  To disentangle size from
cohesive effects, we show the rescaled diffusivity $\IDcG/\IDo$ in
\fig~\ref{fig:6}b. For $\Ieps\rightarrow0$ the rescaled diffusivity
approaches unity, indicating that internal friction is
unimportant. The rescaled diffusivity exhibits for small cohesion only
a weak dependence on the size of the globule. As $\Ieps$ approaches
the solid regime, however, deviations between different system sizes
become observable. Thus, we propose that in the liquid regime, the
internal friction is extensive and to leading order scales with the
size of the globule
\begin{equation}
  \label{eq:18}
  \IDcG \sim  1/\INGred 
  \eqspace
\end{equation}
as will be corroborated by our non-equilibrium simulations in
section~\ref{sec:impact-intern-frict}.  Further below we will also
show how the internal viscosity within the globule can be extracted
from the diffusivity ratio $\IDcG/\IDo$.

\section{Forced unfolding of globules: non-equilibrium simulations}
\label{sec:impact-intern-frict}

\subsection{Model}
\label{sec:model-4}
We now discuss our simulations for the non-equilibrium force-induced
dissolution of homopolymeric globules.  Here, the two chain ends are
positionally constraint at $\vecsc{\vec r}{1}{}(\It) =
-\Ipotentialtrapr(\It)$ and $\vecsc{\vec r}{\IN}{}(\It) =
\Ipotentialtrapr(\It)$. The trap positions are moved at constant speed
$\IpotentialtrapPos(\It) = \IpotentialtrapPosmin + \Iv \It$ with
$\IpotentialtrapPosyz = 0$, from $\IpotentialtrapPosmin$ up to a
maximal position $\IpotentialtrapPosmax =\Ia(\IN-1)$, producing a
time-varying chain extension $\Iextension(\It) = 2
\IpotentialtrapPos(\It)$.  During the pulling, the force acting on the
terminal beads is measured.  The backbone bonds are modeled by a
harmonic potential $\Ipotentialbond$ as before, see \eq~\eqref{eq:8}.
Excluded volume and cohesive interactions are again modeled by a
Lennard-Jones potential $ \IpotentialLJ$,
\eq~\eqref{eq:10}. \Eq~\eqref{eq:4} is used to integrate the Langevin
equation. We no longer employ periodic boundary conditions, however
still prevent knot formation by introducing a potential, which mimics
two repulsive bars that extend from the first/last bead to the
left/right along the $x$-axis
\begin{equation}
  \label{eq:19}
  \Ipotential_{\mathrm{k}} =
  \begin{cases}
    \sum_{i=2}^{\IN-1}\left( \left({2\Ia}/{
          \vecsc{\hat\rho}{i}{}}\right)^{12} - 2 \left({2\Ia}/{
          \vecsc{\hat\rho}{i}{}}\right)^{6} +1 \right) & \text{if $
      \vecsc{\hat\rho}{i}{}< 2\Ia$ and
      $|\vecsc{\Ir}{i}{x}|>\IpotentialtrapPos$}\\
    0 & \text{else,}
  \end{cases}
\end{equation}
$\vecsc{\hat\rho}{i}{} =
\sqrt{{\vecsc{\Ir}{i}{y}}^2+{\vecsc{\Ir}{i}{z}}^2}$.
$\Ipotential_{\mathrm{k}}$ does not affect the stretching response.
The total energy is therefore given by $\Ipotential = \Ipotentialbond
+ \IpotentialLJ + \Ipotential_{\mathrm{k}}$.  Two different protocols,
annealed and un-annealed, are used for the initial configurations in
order to investigate the history dependence on the globule pulling
response. We record force extension curves for various chain lengths
$\IN = 50,\,100,\,200,\,300$, cohesive strengths $0\leq\Iepsdl
\leq4.1$, and pulling velocities $\Ivdl= \Iv/(\Ia/\Itd) =
0.001,\,0.0045,\,0.01,\,0.0225,\,0.045$ going significantly beyond our
previous work, where the largest system was $\IN = 100$, the slowest
velocity was $\Ivdl = 0.0045$, and only un-annealed initial
configurations were used~\cite{Alexander-Katz2009}.  For each
parameter set, twenty stretching cycles are simulated.

\begin{figure}
  \centering
  \includegraphics{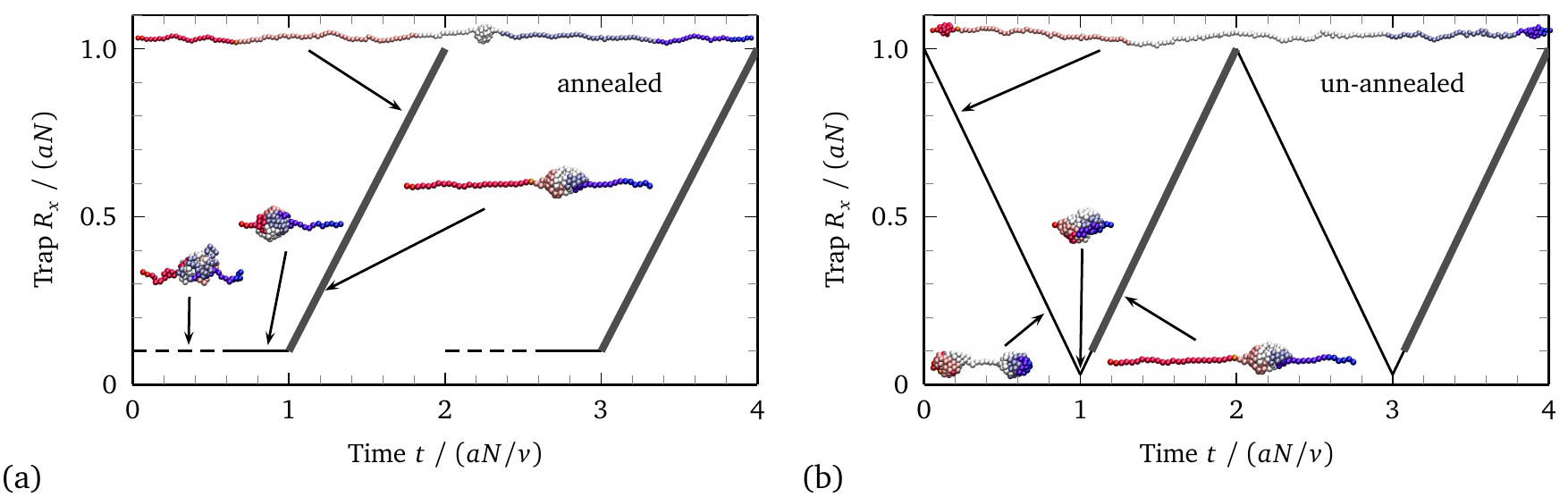}%
  \subfloat{\label{fig:8a}}%
  \subfloat{\label{fig:8b}}%
  \caption{Illustration of the two different pulling protocols.
    $\protect\IpotentialtrapPos$ and $-\protect\IpotentialtrapPos$
    denote the positions of the two traps that positionally constrain
    the chain ends.  \subref{fig:8}a~Preparation of the annealed
    structures. After a long equilibrium run with $\Iepsdl = 0.8$ and
    $\protect\IpotentialtrapPos = 0.1$ (broken line), the globule is
    equilibrated at the cohesive strength at which the pulling curve
    is recorded (horizontal solid line). The subsequent pulling cycle
    is depicted by the thick gray line. \subref{fig:8}b~Preparation of
    the un-annealed structures. The traps are moved from complete
    extension, $\protect\IpotentialtrapPos=a(N-1)$, to
    $\protect\IpotentialtrapPos =0.03\Ia\IN$, which ensures that one
    single globule forms. Without pausing, the pulling cycle starts
    and force extension curves are recorded and analyzed in the
    interval $0.1<\protect\IpotentialtrapPos/(\Ia\IN)<1$ (thick gray
    line). }
  \label{fig:8}
\end{figure}
\begin{figure}
  \centering
  \includegraphics{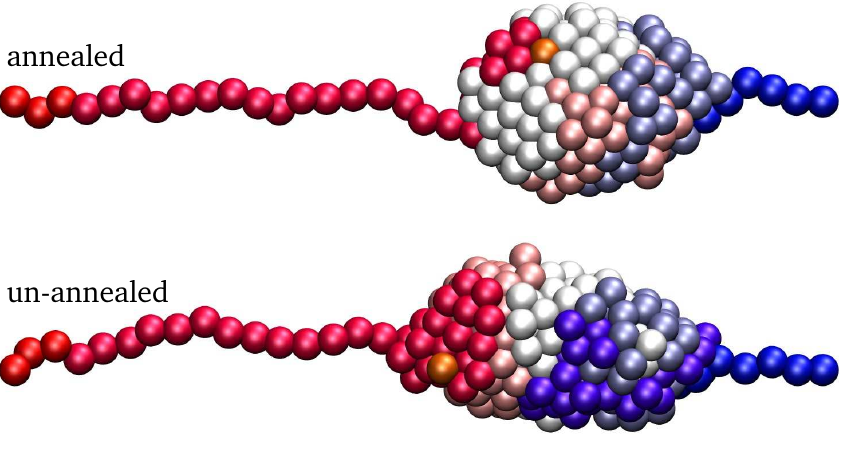}
  \caption{Typical initial configurations in the annealed (top) and
    un-annealed simulations (bottom) for $\Iepsdl = 2.91$ and $\Ivdl =
    0.001$. For the large $\Ieps$ shown here, the collapsed
    un-annealed structures show some residual ordering. The color
    coding indicates the monomer index along the chain contour.}
  \label{fig:9}
\end{figure}

\subsubsection{ Annealed initial structures}
\label{sec:pull-rand-struct}
Annealed initial structures are obtained by performing an equilibrium
annealing simulation with moderate cohesive strength $\Iepsdl = 0.8$
and fixed trap position $\IpotentialtrapPos = 0.1\Ia\IN$. The LJ
interaction is strong enough to induce globule formation, yet small
enough to allow for rapid equilibration of the chain conformation
inside the weakly collapsed globule. Knot formation is prevented by
virtue of the potential in \eq~\eqref{eq:19}.  Every $\It = 20000\Itd
$ a structure is recorded, which is subsequently equilibrated for $\It
= 10000\Itd$ using the target cohesive strength at which the pulling
simulation is to be conducted. The resulting structure is used as one
initial configuration for the subsequent pulling cycle. In
\fig~\ref{fig:8}a the pulling protocol is illustrated and in
\fig~\ref{fig:9} typical initial configurations are depicted.

\subsubsection{ Un-annealed initial structures}
\label{sec:pull-order-struct}
This set of initial structures is obtained by starting from an
extended configuration and moving the traps from $
\IpotentialtrapPosmax = \Ia(\IN-1)$ to $\IpotentialtrapPos
=0.03\Ia\IN$. Typically, we observe the formation of one globule for
$\Ivdl<0.01$ and the formation of two globules near the traps for
$\Ivdl>0.01$, which merge at small extension, see snapshots in
\fig~\ref{fig:8}b.  Without pausing, the traps are subsequently
extended to $ \IpotentialtrapPos = 0.1\Ia\IN$, where the actual
pulling cycle starts and force extension curves are recorded for
further analysis. During the compression stage, the traps are moved
with the same velocity $\Iv$ with which the force extension curve is
recorded, \fig~\ref{fig:8}b. For larger cohesive strengths~$\Ieps$,
non-equilibrium ordered structures prevail as initial configurations
for the subsequent pulling cycle, see \fig~\ref{fig:9} for an
illustration. Although not the prime target of our present
investigation, such structures might be of relevance when studying the
dynamics and packing of DNA chromatin structures far from
equilibrium~\cite{Grosberg1988,Grosberg1993,Lieberman-Aiden2009}.

\subsection{Liquid-solid transition for large cohesive strengths}
\label{sec:glass-trans-visible}

\begin{figure}
  \centering
  \includegraphics{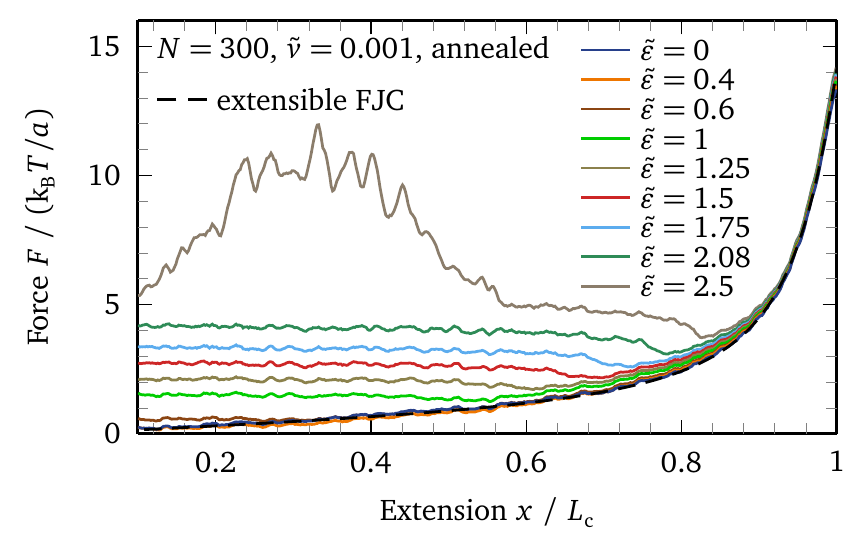}
  \caption{Force extension curves for constant velocity $\Ivdl =
    0.001$, various cohesive strengths and $\IN = 300$ using annealed
    initial configurations. All curves are averages over twenty
    pulling cycles. Above the globule transition,
    $\Iepsdl>\IepsGdl\approx0.5$, a force plateau followed by a dip in
    the force extension curve is observed. The broken line depicts the
    theoretically expected force extension trace of an extensible
    freely jointed chain for $\Iepsdl = 0$, \eq~\eqref{eq:20}. For
    $\Iepsdl = 2.5$ the pulling curve exhibits a marked maximum at
    small extensions, which indicates frozen internal dynamics. }
  \label{fig:10}
\end{figure}

In \fig~\ref{fig:10}, stretching curves averaged over twenty pulling
cycles for the annealed pulling protocol are shown for various
cohesive strengths $\Ieps$ and $\Ivdl = 0.001$, $\IN = 300$.  Beyond
the globule transition,
$\Iepsdl>\IepsGdl\approx0.5$~\cite{Alexander-Katz2006,Alexander-Katz2009,Parsons2006},
an $\Ieps$~dependent force plateau is observed. The plateau
force~$\IFp$ increases as $\Ieps$ increases and is, for the relatively
slow pulling speed shown here, mostly associated with the equilibrium
free energy per unit length of globule formation, $\IFp\Ia\sim
\Ieps-\IepsG$~\cite{Alexander-Katz2009}.  For large extensions of the
order of $\Iextension/\Icontourlength \approx0.8$, a dip in the force
extension curve appears, which is the signature of a pulling induced
globule dissolution. For even larger extension, essentially no monomer
contacts are present, and the force extension curve becomes
independent of the cohesive strength $\Ieps$ and follows the trace of
an extensible freely jointed chain (shown as a broken line)
\begin{equation}
  \label{eq:20}
  \Iextension / \Icontourlength = \coth(2\Ia\IF/(\kBT)) + \kBT/(2\Ia\IF) +
  \IF/(2\Ia\IpotentialbondK)
  \eqspace.
\end{equation}
A phantom chain ($\Iepsdl = 0$) coincides perfectly with
\eq~\eqref{eq:20}.  The curve for $\Iepsdl = 2.5$ features a maximum
at $\Iextension/\Icontourlength\approx0.3$, whose origin will be
discussed in the next paragraph.

\begin{figure}
  \centering
  \includegraphics{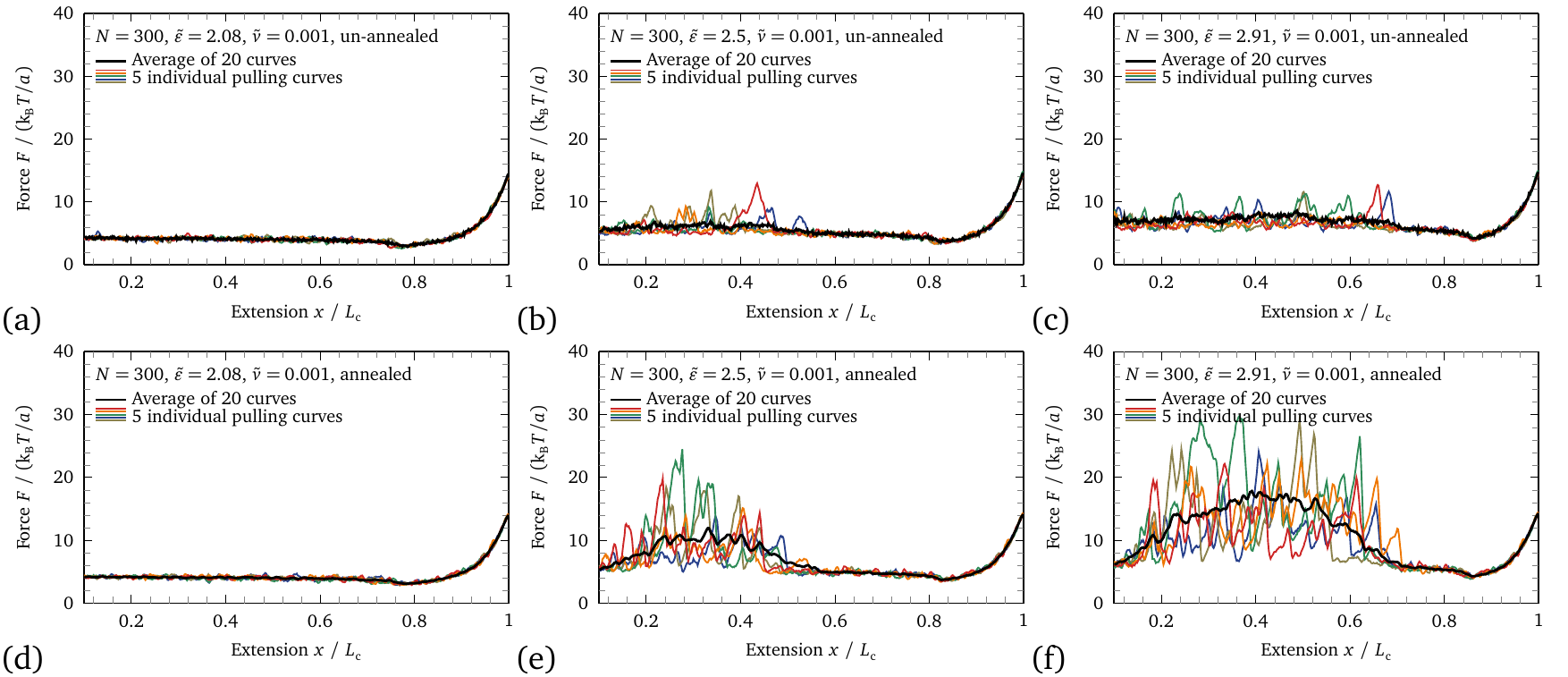}
  \caption{Pulling curves for $\IN = 300$ and $\Ivdl = 0.001$ with
    (a-c)~un-annealed and (d-f)~annealed structures as initial
    configurations.  The thin lines are individual force extension
    traces, whereas the thick black line is the average over twenty
    pulling curves.  The un-annealed structures exhibit rather smooth
    pulling curves and no drastic differences between the various
    cohesive strengths $\Ieps$ is observed.  In contrast, the annealed
    structures feature marked fluctuations above a certain
    threshold~$\Iepscdl\approx2.1$ for $\IN=300$, which diminish for
    large extension, \ie when the globule sizes decrease. We attribute
    these fluctuations to a liquid-solid transition, which
    dramatically changes the internal dynamics (see text). }
  \label{fig:11}
\end{figure}

\Fig~\ref{fig:11} shows force extension traces for the two different
pulling protocols for $\IN = 300$, $\Ivdl = 0.001$, and relatively
strong cohesive forces $\Iepsdl = 2.08,\,2.5,\,2.91$.  The thick curve
in each plot depicts the average over all 20 pulling curves, while the
thin curves show individual force-extension traces for different
initial configurations. In \mbox{\figs~\ref{fig:11}a-c} pulling curves
starting from un-annealed configurations are shown. Increasing $\Ieps$
leads to an increasing plateau force similar to \fig~\ref{fig:10}. For
$\Iepsdl\geq2.5$ stronger fluctuations of the force are observed, but
the force extension traces are qualitatively similar to the curves for
$\Iepsdl = 2.08$. The situation is vastly different for pulling curves
starting from the annealed initial configurations,
\mbox{\figs~\ref{fig:11}d-f}. Again, for small cohesive strengths
$\Iepsdl\leq2.08$ the pulling curves are smooth and no strong
fluctuations occur, \cf \fig~\ref{fig:10}. However, increasing the
cohesive strength further leads to pronounced fluctuations in the
force extension curves. This is due to a transition of the internal
dynamics from liquid-like to
solid-like~\cite{Alexander-Katz2009,Sing2010,Vitkup2000,Taylor2009,Rostiashvili2001,Paul2007,Liang2000}.
Since the un-annealed initial configurations are rather ordered
--~especially for large $\Ieps$~-- the globules are easily unwound by
simply retracing the configurational intermediates that were
encountered upon folding in reversed order.  The liquid-solid
transition occurs for the un-annealed structures, too, but has almost
no effect on the non-equilibrium pulling simulations.  This is very
different for the annealed simulation protocol, where the mean force
is higher and the variation of individual force curves around the mean
force is also pronounced.

\begin{figure}
  \centering
  \includegraphics{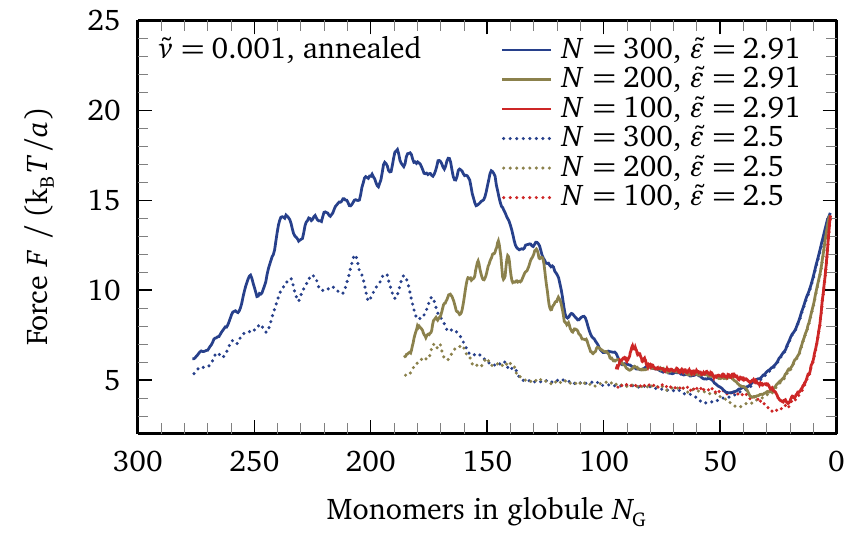}
  \caption{ Averaged force extension curves in the vicinity of the
    liquid-solid transition are shown as a function of the number of
    monomers inside the globule, $\ING$, \eq~\eqref{eq:12}, for the
    annealed set of initial configurations.  Once $\ING$ is below a
    certain threshold ($\approx100$ for $\Iepsdl = 2.91$, $\approx140$
    for $\Iepsdl = 2.5$) the globule is driven into the liquid state,
    and force curves for different $\IN$ and equal $\Ieps$ collapse.
    The dependence of the liquid-solid transition on $\ING$ is
    discussed in detail in ref.~\cite{Sing2010}.}
  \label{fig:12}
\end{figure}

The dependence of the liquid-solid transition on the globule size
$\ING$ is illustrated in \fig~\ref{fig:12}. Here, force curves
obtained using the annealed pulling protocol are plotted versus the
number of monomers~$\ING$ inside the globule for different monomer
numbers $\IN$ and cohesive strengths.  $\ING$ is calculated \via a
modified version of \eq~\eqref{eq:12} where the extension
$\Iextension$ of the polymer is used instead of $\Iboxsize$. One
notices that the pronounced noise in the curves ceases once the
globule is below a certain size, i.e. once $\ING$ is below a certain
threshold value, where this critical size depends on $\Ieps$. This
feature is independent of the chain length~$\IN$ but solely depends on
$\ING$, i.e.  curves with equal $\Ieps$ but different $\IN$ coincide
once the globule size has fallen under the critical
size~\cite{Sing2010,Parsons2006,Parsons2006a,Paul2007,Rampf2005,Zhou1996,Taylor2009,Rostiashvili2001,Liang2000}.
The cohesive strength $\Iepsc$ at which this liquid-solid transition
occurs is also independent of the pulling velocity within the range of
velocities studied. In \fig~\ref{fig:13} we see that for $\Iepsdl =
2.08$ even with the highest pulling velocity, no large fluctuations in
the force extension curves are induced and the globule remains in the
liquid phase. Therefore, the liquid-solid transition at $\Iepsc$ is
not a mere non-equilibrium pulling feature but an indication of a
change in the equilibrium internal dynamics.  The transition observed
in the non-equilibrium pulling simulations is of course the same
transition that induces the abrupt change of the equilibrium globule
diffusivity, demonstrated in \fig~\ref{fig:6}, since both the
equilibrium globule diffusion and the non-equilibrium pulling response
are ultimately related to the configurational chain dynamics inside
the globule.


\subsection{Internal friction}
\label{sec:internal-friction}

\subsubsection{Internal friction is viscous }
\label{sec:diss-energy-scal-1}

\begin{figure}
  \centering
  \includegraphics{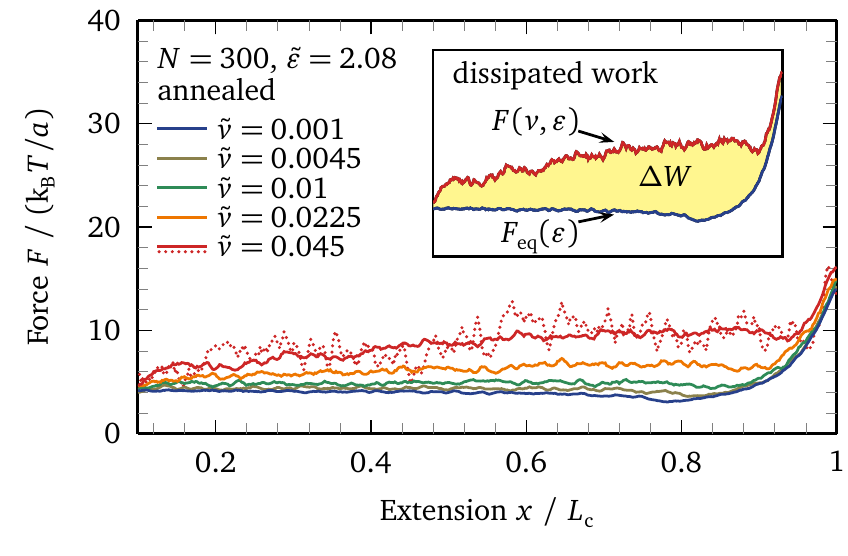}
  \caption{Averaged pulling curves for various pulling velocities with
    $\IN = 300$, $\Iepsdl = 2.08$, and annealed initial
    configurations. The friction force increases with increasing
    pulling velocity. The curves with the slowest pulling velocities
    $\Ivdl = 0.001,\,0.0045$ almost coincide, indicating that with
    these slow velocities one has approximately reached the
    equilibrium pulling limit. The dotted curve is one individual
    pulling curve for $\Ivdl = 0.045$. The inset illustrates the
    definition of the dissipated work, \eq~\eqref{eq:21}, which is the
    shaded area between the two curves.}
  \label{fig:13}
\end{figure}

\begin{figure}
  \centering
  \subfloat{\label{fig:115a}}%
  \subfloat{\label{fig:115b}}%
  \includegraphics{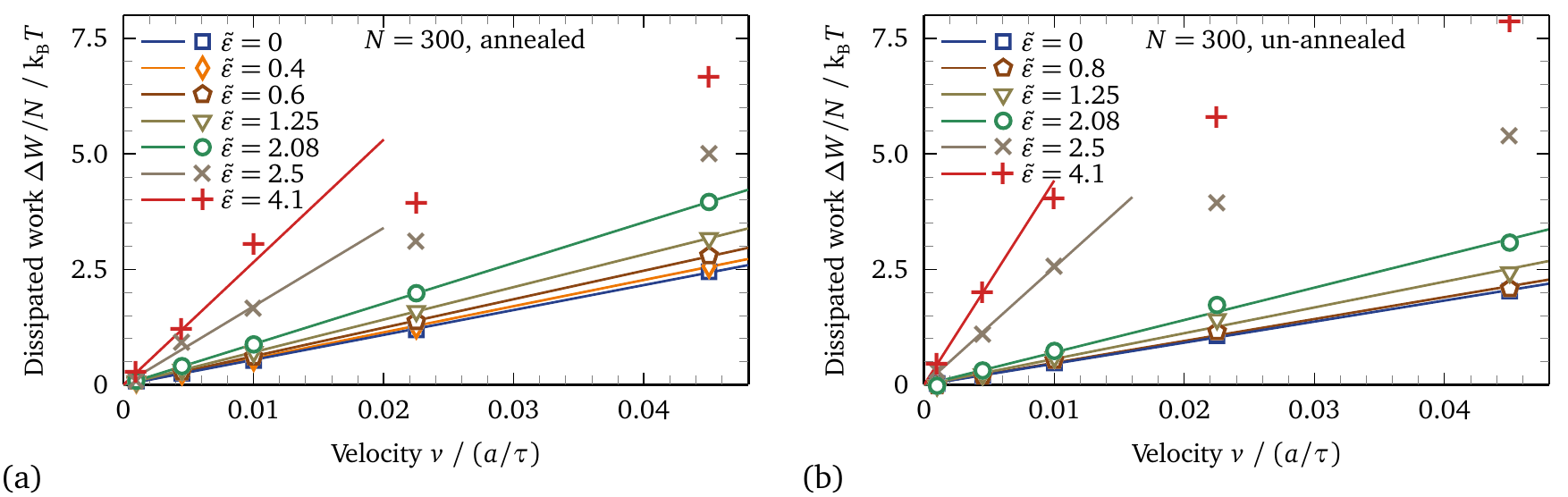}%
  \caption{Dissipated work per monomer $\IWdiss{}/\IN$ as a function of the pulling velocity
    $\Iv$ for $\IN = 300$ and different cohesive strengths~$\Ieps$ starting from the
    (a)~annealed and the (b)~un-annealed set of initial configurations.  Symbols depict
    simulation data, lines show linear fits to the data according to  \eq~\eqref{eq:26}.  Below the globule transition,
    $\Iepsdl<\IepsGdl\approx0.5$, the curves collapse, \ie  interaction contributions to friction are  
negligible.  Below the liquid-solid transition, $\Iepsdl<\Iepscdl\approx2.1$ for $\IN = 300$, 
the data is linear in $\Iv$ in the whole range of velocities studied, indicating that 
friction is of viscous nature.
  Above the liquid-solid transition the linear scaling breaks down.
  $\IWdiss$ is slightly lower for the simulations starting from the
  un-annealed set of initial configurations showing that the globule
  is not completely equilibrated and still rather ordered.  }
\label{fig:115}
\end{figure}

\begin{figure}
  \centering
  \subfloat{\label{fig:14a}}%
  \subfloat{\label{fig:14b}}%
  \includegraphics{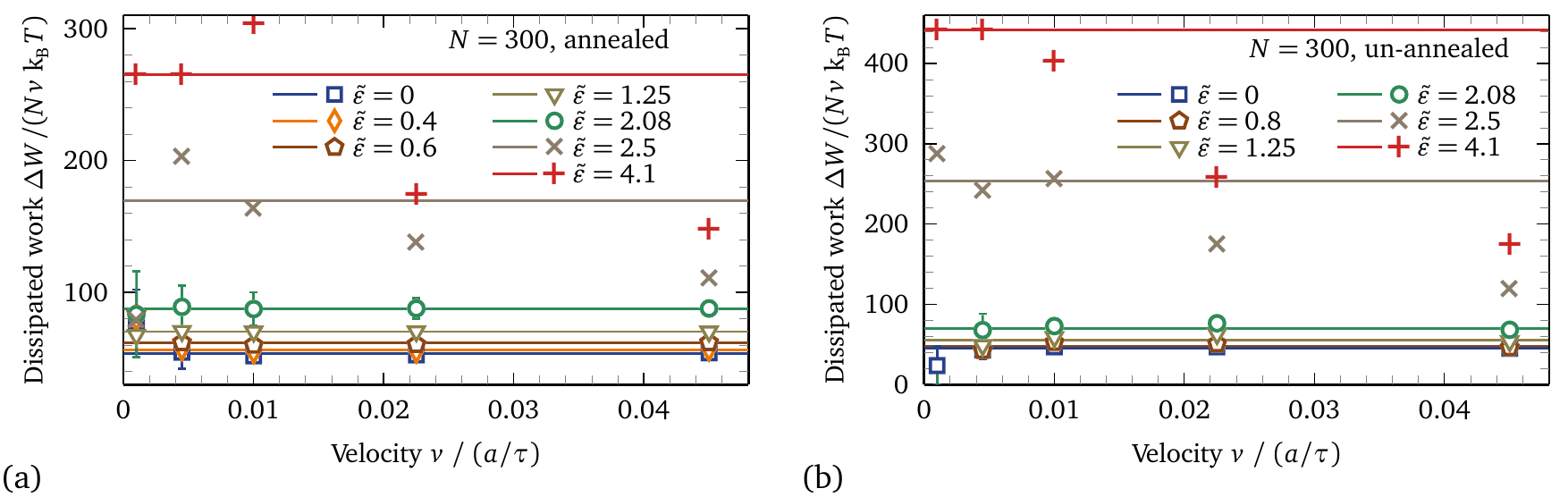}%
  \caption{Dissipated work per monomer rescaled by the  pulling velocity, $\IWdiss{}/(\IN\Iv)$, as
    a function of the pulling velocity $\Iv$ for $\IN = 300$ and different cohesive
    strengths~$\Ieps$ starting from the \subref{fig:14}a~annealed and the
    \subref{fig:14}b~un-annealed set of initial configurations. Symbols denote simulation
    results, lines depict $\Ifrictioncoefficientwork$ as obtained from linear fits according to
    \eq~\eqref{eq:26} in  \fig~\ref{fig:115}. For $\Ieps<\Iepsc$, our simulations
    are carried out in the linear viscous  regime and there is no $\Iv$ dependence after
    rescaling. Above the liquid-solid transition 
    the linear scaling in $\Iv$ breaks down. 
    For $\Iepsdl=0$ and $2.08$, error bars indicate the standard deviations of the twenty
    measurements of $\IWdiss(\Iv,\Ieps)$.  
  }
  \label{fig:14}
\end{figure}

For small $\Ieps<\Iepsc$, the internal chain dynamics is fast and no
significant difference between the simulations with annealed and
un-annealed initial configurations is observed.  In \fig~\ref{fig:13}
pulling curves for various pulling velocities~$\Iv$ are shown. We
observe that even for a cohesive strength as large as $\Iepsdl = 2.08$
and polymers as long as $\IN=300$, the two slowest pulling curves
$\Ivdl = 0.001,\,0.0045$ almost coincide. This indicates that force
extension curves for $\Ivdl = 0.001$ are already very good
approximations for equilibrium curves. Increasing velocity leads to
increasing energy dissipation. There are two major mechanisms leading
to dissipation: solvent friction and internal friction. These
dissipation mechanisms dominate in different parts of the pulling
curve. For small extensions, when most of the monomers are part of the
globule, internal friction dominates. Towards the end of the pulling
process, the globule is markedly smaller and the solvent friction,
which acts mostly on the linker chain, dominates the pulling curve.
The \Tdefinition{dissipated work} $\IWdiss(\Iv, \Ieps)$ is defined as
the difference between the total work at finite velocity and the
equilibrium work~\cite{Alexander-Katz2009}
\begin{equation}
  \label{eq:21}
  \IWdiss(\Iv, \Ieps) = \IW(\Iv, \Ieps)  - \IWeq(\Ieps)
  \eqspace,
\end{equation}
where the \Tdefinition{work} done by one trap is generally defined as
\begin{equation}
  \label{eq:22}
  \IW =  - \int_{ \IpotentialtrapPosmin}^{ \IpotentialtrapPosmax} \IF(x)\drm \Iextension
  \eqspace,
\end{equation}
see the inset of \fig~\ref{fig:13} for an illustration of this
definition.  The \Tdefinition{equilibrium work}~$\IWeq(\Ieps)$ is
obtained from extrapolating $\IW(\Iv, \Ieps)$ to $\Iv\rightarrow0$.
In \fig~\ref{fig:115} we show the dissipated work per monomer,
$\IWdiss /\IN$, as a function of the velocity~$\Iv$ for $\IN = 300$
and various $\Ieps$. Below the liquid-solid transition,
$\Iepsdl<\Iepscdl\approx2.1$ for $\IN = 300$, the dissipated work
scales linearly with $\Iv$, which is further demonstrated in
\fig~\ref{fig:14}, where the ratio $\IWdiss/ ( \Iv \IN) $ is
plotted. This shows that in the liquid state, the simulations are
conducted in the experimentally relevant linear response regime and
the friction is essentially of viscous nature.  Above the liquid-solid
transition this scaling breaks down, or, in other words, the
velocities probed in the simulations are not low enough to reach the
viscous regime. Further, we observe that below the globule transition,
$\Iepsdl<\IepsGdl\approx0.5$, the dissipated work is almost
independent of the cohesive strength.
This suggests that monomer-monomer attraction and topological
constraints, \eg entanglements, are negligible for small values of the
cohesive strength, i.e. in the non-collapsed state.  In order to
extract such subtle friction effects, much longer simulations would be
needed.
Knots, which in principle might arise due to bond crossing, are not
observed in our simulations.

\begin{figure}
  \centering
  \includegraphics{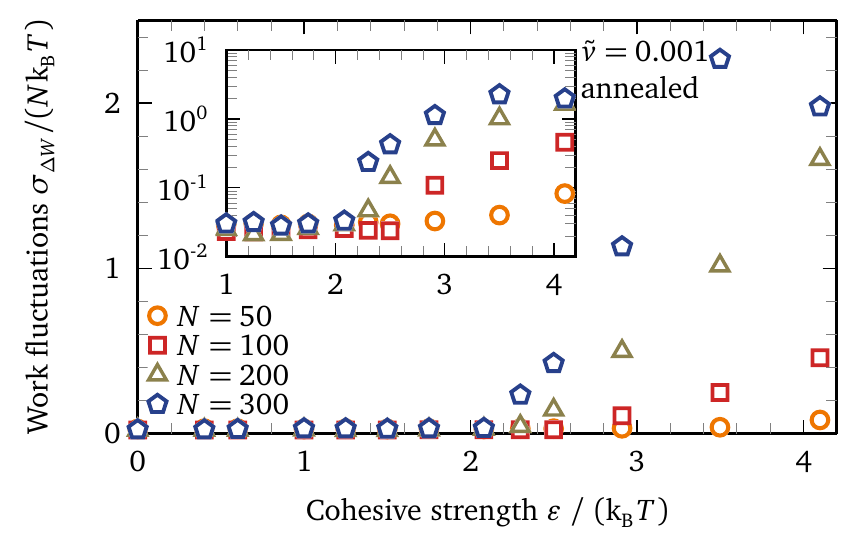}
  \caption{The standard deviation per monomer
    $\standarddeviation{\IWdiss}/\IN$ of the dissipated work, as
    obtained from twenty pulling cycles.  For $\Ieps<\Iepsc$,
    $\standarddeviation{\IWdiss}/\IN$ is independent of the chain
    length~$\IN$ and small. The fluctuations of the dissipated work
    increase significantly when $\Ieps>\Iepsc$, where $\Iepsc$
    decreases for increasing $\IN$, see \tab~\ref{tab:1}. The data is
    obtained from simulations with $\Ivdl = 0.001$ and annealed
    initial configurations. }
  \label{fig:15}
\end{figure}

The fluctuations of the dissipated work can be conveniently used to
define the liquid-solid transition. In \fig~\ref{fig:15} the standard
deviation
\begin{equation}
  \label{eq:29}
  \standarddeviation{\IWdiss}  = \sqrt{k/(k-1)} \timesspace \sqrt{\expectation{\IWdiss^2-\expectation{\IWdiss}^2}}
\end{equation}
obtained from $k=20$ measurements of $\IWdiss$ for $\Ivdl = 0.001$ in
the annealed pulling protocol is plotted against $\Ieps$ on both
linear and logarithmic scales.  For $\Ieps<\Iepsc$,
$\standarddeviation{\IWdiss}/\IN$ is independent of $\IN$ and
$\Ieps$. As one crosses from the liquid into the solid phase at
$\Ieps=\Iepsc$, the fluctuations increase. \Fig~\ref{fig:15} clearly
demonstrates that $\Iepsc$ depends on the system size $\IN$. The
estimates of $\Iepsc$ extracted from the behavior of
$\standarddeviation{\IWdiss}/\IN$ are shown in \tab~\ref{tab:1}.
\begin{table}
  \centering
  \begin{tabular}{l|cccc}
    $\IN$&100&150&200&300\\\hline
    $\Iepscdl$&4&2.9&2.3&2.1
  \end{tabular}
  \caption{The critical cohesion strength~$\Iepsc$ of the liquid-solid transition depends on the length
    of the polymer~$\IN$. For $\Ieps<\Iepsc$, the system is in the liquid phase and fluctuations
    of the dissipated work are small. For $\Ieps>\Iepsc$ pronounced fluctuations are observed
    due to slow internal dynamics, see \fig~\ref{fig:15}. }
  \label{tab:1}
\end{table}

\subsubsection{Internal friction is extensive}
\label{sec:diss-energy-scal}
We demonstrated that in the liquid state the dissipated work~$\IWdiss$
and thus the internal friction scale linearly with the
velocity~$\Iv$. Therefore, we can express the friction
force~$\IFfriction$ in terms of a velocity independent viscous
friction coefficient~$\Ifrictioncoefficient(\Ieps,\ING)$, which
depends on the number of monomers in the globule,
\begin{equation}
  \label{eq:23}
  \IFfriction(\Iv,\Ieps,\ING) = \IF(\Iv,\Ieps,\ING) - \IFeq(\Ieps,\ING)
  = \Iv\timesspace  \Ifrictioncoefficient(\Ieps,\ING)
  \eqspace,
\end{equation}
where $\IFeq$ is the equilibrium force extension curve.  Analogous to
the Stokes friction of a sphere, we define the friction coefficient
as~\cite{Alexander-Katz2009}
\begin{equation}
  \label{eq:24}
  \Ifrictioncoefficient(\Ieps,\ING)\sim  \Ia \IetaG(\Ieps)   \ING^{\Ifrictionexponent}
  \eqspace,
\end{equation}
where $\IetaG(\Ieps)$ is the internal viscosity, which depends on
$\Ieps$ but not on $\ING$. The exponent $\Ifrictionexponent$ describes
the dependence on the monomer number inside the globule, and
characterizes the friction mechanism at work during unraveling the
globule. Two limits can be distinguished: first, if
$\Ifrictionexponent=0$, the friction force is independent of the
globule size~$\ING$, and only a finite number of monomers, which does
not scale with $\ING$, contribute to dissipation. We call this limit
local or {\em intensive friction}, and the pictorial mechanism for
such a scenario could be that monomers are peeled off one-by-one from
the surface of the globule.
Second, $\Ifrictionexponent = 1$ describes the limiting situation
where a finite fraction of the globule that is proportional to $\ING$,
or even the entire chain, rearranges in the unfolding process and
hence contributes to the friction force.  \trerevision{We call this
  limit global or {\em extensive friction} and note that there are
  several conceivable microscopic mechanisms.  One possibility could
  be a reptation scenario, where a reorganization of the globule
  conformation is accompanied by a reptation of the entire globular
  chain section (or a chain fraction of length proportional to $\ING$)
  through the globule.  Another possibility would be that the stress
  propagates through the whole globule and leads to dissipation
  without an actual reptation of the chain. We did not analyze the
  precise microscopic dissipation mechanism and therefore leave this
  issue for future work.}

\begin{figure}
  \centering
  \includegraphics{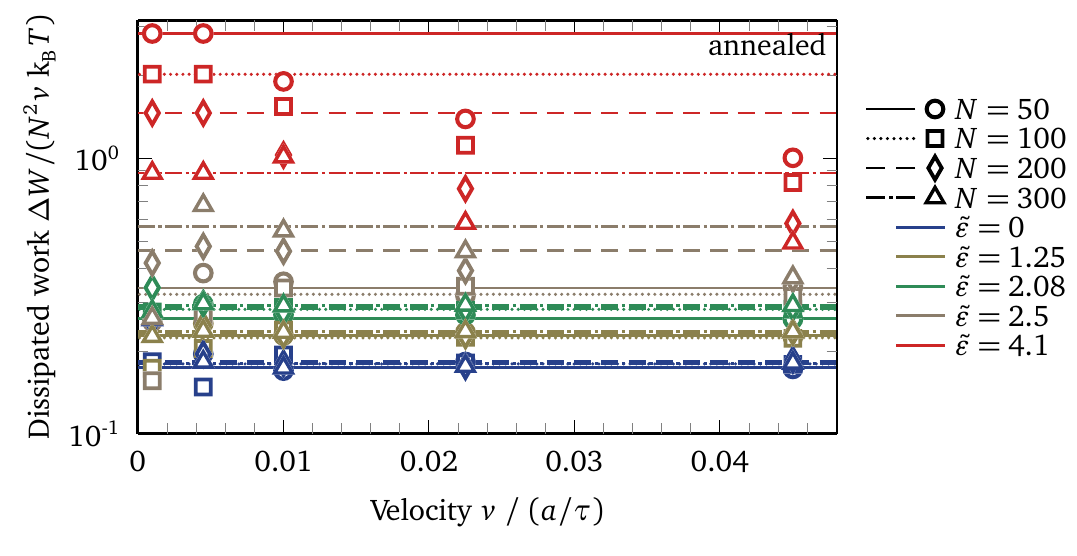}
  \caption{Rescaled dissipated work per monomer $\IWdiss{} /
    (\IN^2\Iv)$ as a function of the pulling velocity $\Iv$ for
    different $\IN$ and cohesive strengths~$\Ieps$ starting from the
    annealed set of initial configurations. Curves with equal $\Ieps$
    and different $\IN$ collapse for $\Ieps<\Iepsc$. This shows that
    the dissipated work per monomer is an extensive function and
    $\IWdiss\sim\IN^2\Iv$. Again, this scaling behavior breaks down
    above the liquid-solid transition.}
  \label{fig:16}
\end{figure}

Integrating the friction force, \eq~\eqref{eq:23}, yields the scaling
form of the dissipated work
\begin{equation}
  \label{eq:25}
  \IWdiss(\Iv,\Ieps,N) = -\int\IFfriction(\Iv,\Ieps,\ING)\drm \Iextension
  \sim  \IetaG( \Ieps) \Ia^2 \IN^{\Ifrictionexponent+1} \Iv
  \eqspace{},
\end{equation}
where we assume $\ING\approx(\Icontourlength - x)/(2\Ia)$.  In
\fig~\ref{fig:16} $\IWdiss/\IN^2$ is plotted versus $\Iv$, the data
with equal $\Ieps<\Iepsc$ and different $\IN$ collapse.  This implies
that the dissipated energy per monomer $\IWdiss/\IN$ and also the
internal friction is extensive, $\Ifrictionexponent = 1$, meaning that
a finite fraction of the globule or the whole globule rearranges
during pulling and not only a few monomers.  \trerevision{This scaling
  behavior breaks down above the liquid-solid transition, which
  however does not imply that the friction becomes non-extensive but
  rather that linear viscous scaling is not valid anymore.  It is also
  not clear whether the extensive scaling of the viscosity, defined
  here via the viscous force when pulling a chain segment out of the
  globule, should be expected to persist in the limit of very large
  globules, as the scaling suggests the viscosity in fact to diverge
  in this thermodynamic limit.  More work on this is also needed.}

\subsubsection{Internal viscosity}
\label{sec:internal-viscosity}
\begin{figure}
  \centering
  \subfloat{\label{fig:17a}}%
  \subfloat{\label{fig:17b}}%
  \subfloat{\label{fig:17c}}%
  \subfloat{\label{fig:17d}}%
  \includegraphics{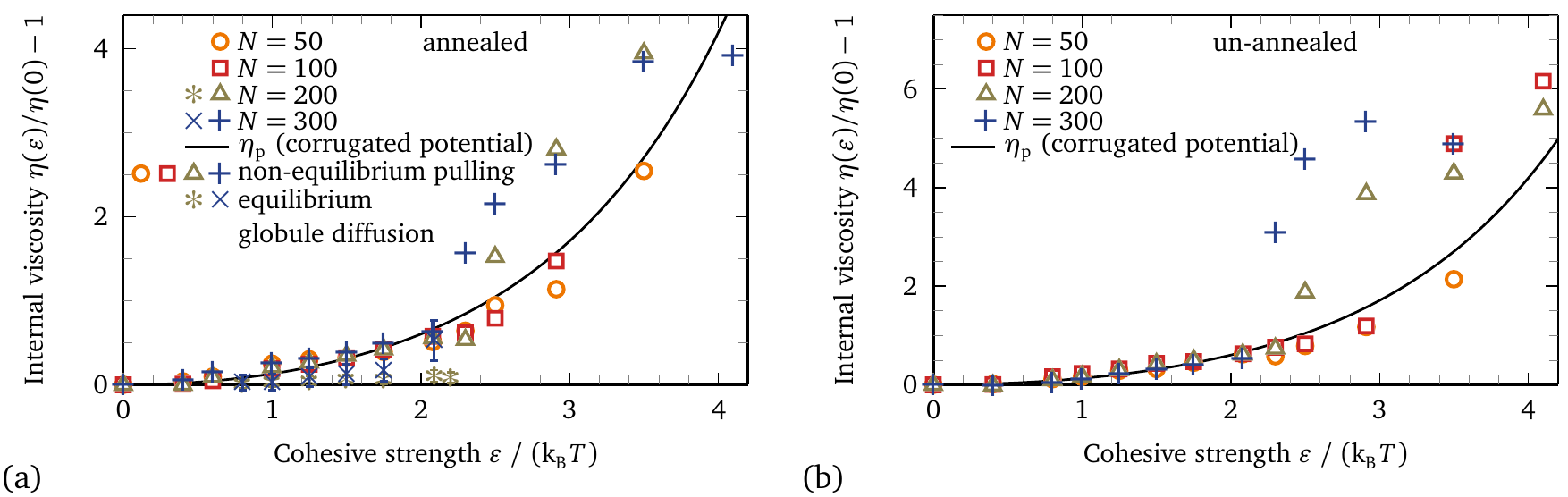}\\%
  \includegraphics{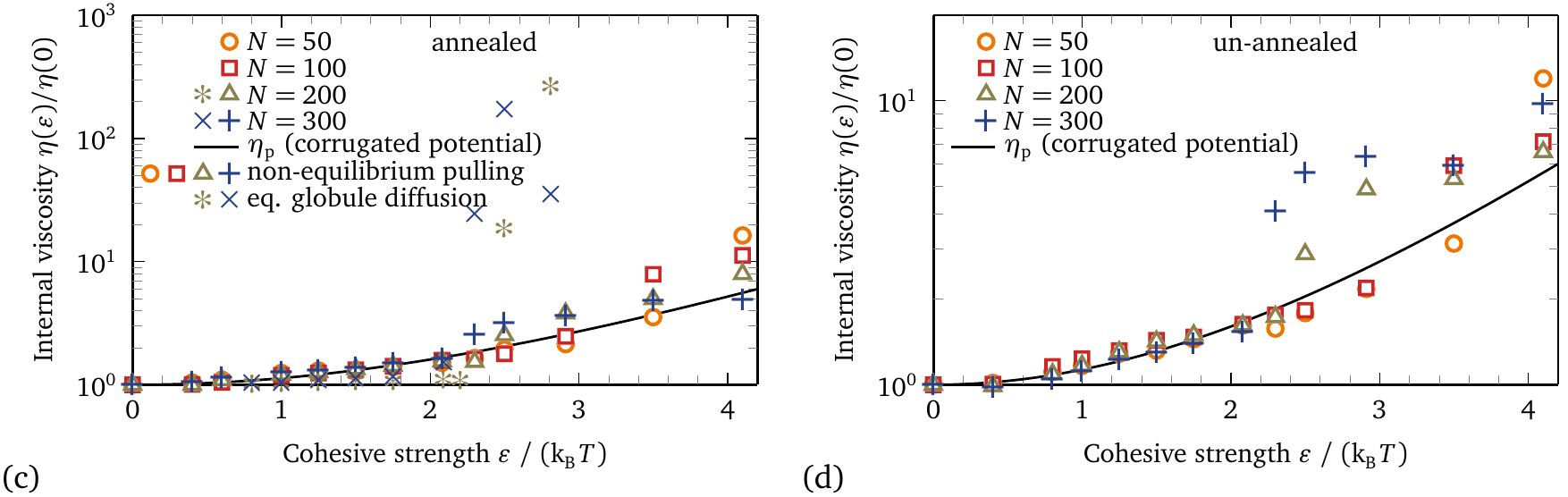}%
 \caption{ Rescaled internal  viscosity $\IetaG(\Ieps)/\IetaG(0) $, \eq~\eqref{eq:27}, as a
    function of the cohesive strength~$\Ieps$ for \subref{fig:17}a,~\subref{fig:17}c~annealed
    \subref{fig:17}b,~\subref{fig:17}d~un-annealed initial configurations. The viscosity $\IetaG(\Ieps)$ is obtained from
    linear fits to the dissipated work as a function of the velocity~$\Iv$.  As the data points
    for different $\IN$ coincide, all
    $\Ifrictioncoefficientwork(\Ieps,\IN)$ exhibit the same $\IN$~dependence.
    The solid line is the prediction of the excess viscosity of a
    Brownian particle in a sinusoidal potential, see
    \eq~\eqref{eq:28}. In \subref{fig:17}a,~\subref{fig:17}c~the
    viscosity ratio as obtained from our equilibrium simulations, see
    \fig~\ref{fig:6}b, ${\IetaG(\Ieps)}/{\IetaG(0)} =
    \IDo/\IDcG(\Ieps) $ is shown in addition.}
  \label{fig:17}
\end{figure}

To extract the internal viscosity quantitatively, we fit the
dissipated work to a linear form in $\Iv$ according to
\begin{equation}
  \label{eq:26}
  \IWdiss(\Iv,\Ieps,\IN) = \Ifrictioncoefficientwork(\Ieps,\IN)\timesspace \Iv
  \eqspace,
\end{equation}
and extract the prefactor $\Ifrictioncoefficientwork(\Ieps,\IN) $,
which is apart from geometrical prefactors proportional to the
friction coefficient $\Gamma$ defined in \eq~\eqref{eq:23}.  As can be
seen in \fig~\ref{fig:115}, $\IWdiss\sim \Iv$ for all velocities
studied if $\Ieps<\Iepsc$. Consequently we include all velocity data
in the linear fit.  However, for $\Ieps>\Iepsc$ a marked deviation
from linear behavior is observed. There, we fit only to the slowest
velocities, where the data still scales linearly with $\Iv$.  From the
scaling $\Ifrictioncoefficientwork(\Ieps,\IN) \sim \IetaG(\Ieps)
\IN^{\Ifrictionexponent + 1}$ one sees that the relative internal
viscosity follows as
\begin{equation}
  \label{eq:27}
  \frac{\IetaG(\Ieps)}{\IetaG(0)}  =  \frac{\Ifrictioncoefficientwork(\Ieps,\IN)}{\Ifrictioncoefficientwork(0,\IN)}
  \eqspace,
\end{equation}
which is shown in \fig~\ref{fig:17} for annealed and un-annealed
initial configurations.  Note that in the limit of vanishing cohesion,
$\Ieps=0$, the globular viscosity $ \IetaG(\Ieps) $ is due to solvent
friction effects only. The internal globular viscosity without solvent
effects can therefore be defined by the difference $ \IetaG(\Ieps) -
\IetaG(0) $, but note that this definition is approximate since the
solvent friction also depends weakly on the value of $\Ieps$ since the
globular structure changes with varying cohesive strength.  By
plotting the ratio of the viscosity in the presence and absence of
cohesion, any residual numerical prefactors and the polymer length
dependence in \eq ~\eqref{eq:25} are eliminated and we are able to
compare all different sets of parameters.  The internal viscosities
extracted from the non-equilibrium pulling simulations coincide for
different~$\IN$, which indicates that the scaling Ansatz
$\Ifrictioncoefficientwork(\Ieps,\IN) \sim \IetaG(\Ieps)
\IN^{\Ifrictionexponent + 1}$ works fine.  In \fig~\ref{fig:17}a for
the annealed pulling simulations we add data for the internal
viscosity obtained from the equilibrium globule diffusion simulations,
which are defined via ${\IetaG(\Ieps)}/{\IetaG(0)} =
\IDo/\IDcG(\Ieps)$ (the same data as already presented in
\fig~\ref{fig:6}b). The viscosity data from the equilibrium
simulations are considerably lower for $\Ieps<\Iepsc$. This might in
part be caused by an underestimate of the Rouse friction $\IDo$
defined in \eq~\eqref{eq:16}, which enters the definition of the
viscosity in the equilibrium simulations, or by additional dissipative
mechanisms in the non-equilibrium pulling simulations: sometimes it
happens that the whole globule is moved through the solvent when an
entanglement within the globule does not yield quickly enough, which
causes additional solvent dissipation.  Considering that the two ways
of extracting the internal viscosity are very different, in terms of
the geometry employed and the general setup (one being equilibrium,
the other non-equilibrium), we consider the agreement sufficient at
the present stage.

In \fig~\ref{fig:17} we compare our data to a simple model based on
the effective friction coefficient of a single particle diffusing in
the periodic potential $\Ipotentialperiodic = ( \Iepsscale\Ieps/2 )
\cos(\pi x/ \Ia)$ with amplitude $\Iepsscale\Ieps$.  This potential
mimics the energy landscape the chain monomers are experiencing as
they are moving against each other during conformational
reorganization processes.
In writing the amplitude of the potential as $\Iepsscale \Ieps$, we
assume the corrugation strength to be proportional to the cohesive
energy $ \Ieps$. The numerical prefactor $\Iepsscale$ is a scaling
factor that may be viewed as a fitting parameter.
The solution of this one-dimensional diffusion problem yields an
effective viscosity~\cite{Zwanzig1988}
\begin{equation}
  \label{eq:28}
  \frac{\Ieta(\Ieps)}{\Ieta(0)} = \BesselIpow{0}{\frac{\Iepsscale\Ieps}{2\kBT}}{2}
\end{equation}
in the stationary long-time limit. $\BesselI{0}{z}$ is the zeroth
order modified Bessel function with the asymptotic limits
$\BesselI{0}{z}\sim1+z^2/4$ for $z\ll1$ and $\BesselI{0}{z}\sim
\e^{z}/\sqrt{2\pi z}$ for $z\gg1$~\cite{Abramowitz2002}.  As can be
seen from \figs~\ref{fig:17}, the rescaled internal viscosity from
simulations, as defined in \eq~\eqref{eq:27}, is reproduced quite well
by the model prediction \eq~\eqref{eq:28} with a fitting value
$\Iepsscale=1$, which is shown by a solid line.  Our results compare
excellently to our previous results where smaller globules
$\IN\leq100$ have been considered~\cite{Alexander-Katz2009}.

In the solid regime for large cohesive energies $\Ieps>\Iepsc$,
pronounced deviations between the equilibrium diffusional and the
non-equilibrium pulling simulations appear, which are better
appreciated when the whole data set is plotted on a logarithmic scale
in \fig~\ref{fig:17}c.  The pulling simulations give a much smaller
viscosity when compared to the equilibrium diffusion simulations,
which shows that the pulling simulations for $\Ieps>\Iepsc$ do not
reach the linear response regime, a fact that is independently
suggested by the absence of scaling in \fig~\ref{fig:115}.

\subsection{Summary for pulling on homopolymer globules}
\label{sec:summary-pulling-homo}
The dissipated work per monomer scales linearly in the pulling
velocity, indicative of viscous friction; the friction is also
proportional to the number of monomers in the globule and scales
extensive, thus a finite fraction of the globule contributes to the
internal friction, \fig~\ref{fig:16}.  We show that the dependence of
the internal friction on the cohesive strength~$\Ieps$ is described
well by the diffusion of a single particle in a corrugated
potential. We extend our previous results~\cite{Alexander-Katz2009} to
significantly larger systems and are able to show that --~below the
liquid-solid transition~-- the history, \ie the preparation of the
initial structures, does not influence the scaling results, see
\fig~\ref{fig:17}.

\section{Summary and conclusions}
\label{sec:summary-conclusions}

Two different dynamical regimes exist for a globular homopolymer, a
solid-like regime at large cohesive energies or low temperatures,
characterized by a substantially increased force needed to unfold the
globule and by pronounced fluctuations in the force extension traces,
and a liquid-like regime at low cohesive energies or high
temperatures.  The critical cohesive energy ~$\Iepsc$ at the
transition between these two regimes depends on the size of the
globule~$\ING$ , as shown in \figs~\ref{fig:12} and~\ref{fig:15}.
In the liquid regime the monomers inside the globule are rather mobile
and the internal friction or viscosity in the globule can be extracted
from simulations.
The globule dynamics is studied by two different scenarios, (i)~by
considering the equilibrium diffusion of a globule relative to the
linking straight chain sections, and (ii)~by non-equilibrium
stretching simulations.
In both scenarios we find that the internal friction or viscosity is
extensive, thus scaling linearly with $\ING$ and increases with
growing $\Ieps$ until the liquid-solid transition is reached. The
signature of the solid state is a vanishing globule diffusion on the
simulation time scales, $\IDcG = 0$, for the equilibrium globule
diffusion setup, while huge fluctuations in the force extension curves
are seen for the non-equilibrium pulling setup.
The solid state is characterized by very slow internal dynamics and no
reliable estimates for the internal friction can be obtained from our
simulations.

Note that we define the internal viscosity in a rather specialized
way, namely via the viscous force needed to pull a chain segment out
of the globule (other definitions of internal viscosity are possible
and useful in different contexts).  \trerevision{Our prediction that
  the friction of chain motion scales extensively with the globular
  size should be discussed in light of the classical reptation
  scenario~\cite{Gennes1971}: Here, the friction coefficient (per
  monomer) is assumed to be independent of the globule size but rather
  the whole chain is assumed to move or reptate through the melt, also
  giving rise to an extensive scaling of the friction.  This suggests
  that reptation is at the heart of the extensive scaling of the
  viscosity found in our simulations, but other mechanisms are also
  conceivable.  Besides, it is not clear whether the extensive scaling
  of the viscosity holds also in the thermodynamic limit and how a
  possible crossover is brought about. For a melt consisting of
  finite-length polymers, the chain size would constitute a possible
  crossover length, for the hypothetical limit of a single infinitely
  long polymer chain no such crossover length comes easily to mind.}

The most direct experimental realization of our system would be
possible with a homopolymeric globule in a single-molecule setup.  But
there might also be implications for protein folding.  Our results
suggest that a quick collapse of a protein might result in a
kinetically trapped and misfolded hydrophobic core, a so-called molten
globular state, which would take a long time to fold into the native
structure due to high internal friction. Since the friction scales
extensively with the monomer number inside the globular core, the
folding time becomes prohibitively long already for moderate globule
size.  \trerevision{From solvent-viscosity dependent measurements of
  folding times, the internal viscosity of a short $\alpha$-helix
  forming peptide in the absence of solvent viscosity effects has been
  estimated to be of the same order as the solvent viscosity,
  i.e. $\IetaG(\Ieps)/ \IetaG(0) \simeq 2$ in our notation
  \cite{Jas2001}.  The interaction energy we extract from
  \fig~\ref{fig:17} follows to be of the order of $\Ieps \simeq 2.5
  \kBT$, not unrealistic for typical interaction parameters for
  residue-residue contacts.  When speculating about the relevance of
  our results to protein folding, a few cautious remarks are in order:
  First, protein collapse is driven by the hydrophobic effect, which
  is quite involved due to the presence of water solvent, neglected in
  our treatment using an implicit solvent model with Lennard-Jones
  interactions.  We would think that the scaling properties of the
  internal friction and internal viscosity will not change by adding
  explicit solvent, which however should be tested.  Secondly,
  specific interactions between protein residues, sequence effects and
  the existence of a well defined native state will certainly alter
  the friction behavior as well. One lesson that might be learned from
  our work is that even for the relatively simple case of a
  homopolymeric globule, the concepts of internal friction and
  internal viscosity are far from trivial.  }

\section{Acknowledgements}
The authors would like to thank the DFG for financial support \via
grants NE 810/8 and SFB 863 and the Leibniz Rechenzentrum for
providing computing facilities. T.R.E. acknowledges support from the
Elitenetzwerk Bayern within the framework of CompInt,
C.E.S. acknowledges support from the NDSEG fellowship.

\bibliographystyle{epj}
\bibliography{internal_friction}
\end{document}